\documentclass[11pt]{article}

\usepackage{hyperref}

\begin{document}

\setcounter{page}{1}

\pagestyle{plain} \vspace{1cm}

\begin{center}
\Large{\bf DBI inflation with a non-minimally coupled Gauss-Bonnet term }\\
\small \vspace{1cm} {\bf Kourosh
Nozari\footnote{knozari@umz.ac.ir}}\quad and\quad {\bf Narges Rashidi\footnote{n.rashidi@umz.ac.ir}}\\
\vspace{0.5cm} Department of Physics, Faculty of Basic Sciences,
University of Mazandaran,\\
P. O. Box 47416-95447, Babolsar, IRAN
\end{center}

\date{\today}

\begin{abstract}
We study the inflation in a model with a Gauss-Bonnet term which is
non-minimally coupled to a DBI field. We study the spectrum of the
primordial perturbations in details. The non-Gaussianity of this
model is considered and the amplitude of the non-Gaussianity is
studied both in the equilateral and orthogonal configurations. By
taking various functions of the DBI field, inflaton potential and
the Gauss-Bonnet coupling term, we test the model with observational
data and find some constraints on the Gauss-Bonnet coupling
parameter. 
\end{abstract}
\newpage

\section{Introduction}

An inflationary stage in the early time evolution of the universe
can address successfully at least some of the problems of the
standard big bang cosmology such as the flatness, horizon and relics
problems. The simplest inflationary scenario, is the one in which
the universe is dominated with a single scalar field whose potential
energy dominates over the kinetic term
~\cite{Gut81,Lin82,Alb82,Lin90,Lid00a,Lid97,Rio02,Lyt09}. Another
important property of the inflationary paradigm is that, it provides
a causal mechanism for production of density perturbations needed to
seed the formation of structures in the universe. In a single field
inflationary scenario, the dominant mode of the primordial density
fluctuations is predicted to be adiabatic and Gaussian to a very
good approximation ~\cite{Mal03}. But, many inflationary models
predict a level of non-Gaussianity which is detectable by current
experiments ~\cite{Mal03,Bar04,Che10,Fel11a,Fel11b}. Also, any
footprint of the non-Gaussianity in observations carries a large
amount of information on the cosmological dynamics which derived the
inflationary expansion of the Universe~\cite{Bab04a,Che08}. While
the evolution of the primordial fluctuations during inflation is
usually studied within the linear theory, to explore the
non-Gaussianity of the density perturbation one has to go beyond the
linear theory. In fact, one has to study the three-point correlation
function of the scalar perturbations or its Fourier transform called
the bispectrum. For a Gaussian signal, all odd $n$-point correlation
functions are vanishing. Also, the higher even $n$-point correlators
are given in terms of the sums of the products of the two-point
functions. So, in order to look for a departure from Gaussianity we
should look for a non-zero three-point correlation functions. To
study the three-point correlation functions, it is required a
perturbative treatment up to the second
order~\cite{Lyt09,Bar04,Wan03,Lan11}. So, a non-Gaussian
distribution of the perturbations implies non-linearity in the
cosmological perturbations. A three-point correlation function, in
the Fourier space, depends on the three momenta or wave numbers
($k_{1}$ , $k_{2}$ and $k_{3}$). Because of the translation
invariance, these momenta add up to zero ($k_{1}+k_{2}+k_{3}=0$) and
thus form a triangle. Also, the rotational invariance implies that
the three-point correlation function depends on the three
independent scalar products of these momenta (or the shape of the
triangle) ~\cite{Bab04a,Kom05,Cre06,Lig06,Yad07}.

It is believed that Einstein gravity is a low-energy limit of a
quantum theory of gravity. String theory is the leading candidate
for quantum gravity. This theory suggests that to have a ghost-free
action, quadratic curvature corrections to the Einstein-Hilbert
action is proportional to the Gauss-Bonnet term
($R_{abcd}R^{abcd}-4R_{ab}R^{ab}+R^{2}$). Such term plays a
significant role in the early time Universe
dynamics~\cite{Zwi85,Bou85}. However it turns out that this term
makes no contribution in the equation for dimension $<5$. But, any
coupling between a scalar field and the Gauss-Bonnet term makes the
Gauss-Bonnet term effective even in four dimension
\cite{Noj05,Noj07,Guo09,Guo10}. For higher dimensional extensions
see for instance
~\cite{Bro07,Bam07,And07,Noz08,Noz09a,Noz09b,Noz09c}.

There is another proposal in the string theory that has attracted
much attention over the past years. In this proposal (which is based
on the Dirac-Born-Infeld (DBI) action ~\cite{Sil04, Ali04}), the
inflaton field is identified with radial coordinate (position) of a
D3 brane moving in a ``throat" region of a warped compactified space
with a speed limit imposed upon its motion, affected by both its
speed and the warp factor of the throat (often assumed an $AdS_{5}$
throat). The effective action in this model involves a non-standard
kinetic term and a function of the scalar field besides the
potential which is related to the local geometry of the compact
manifold traversed by the D3-brane \cite{Sil04}. An interesting
phenomenological feature of the DBI inflation is that it leads to
the non-Gaussian signatures in the Cosmic Microwave
Background~\cite{Hua06,Che07}.

Based on these preliminaries, in this paper we consider a
Gauss-Bonnet term in the action that is non-minimally coupled to the
DBI field. After obtaining the main equations of the model, we study
cosmological inflation and the primordial perturbations with details
in this setup. Also, the issue of non-Gaussianity of the
perturbations will be considered and the amplitude of the
non-Gaussianity in the equilateral and orthogonal configuration will
be calculated. Finally we perform a comparison between the
inflationary parameters and the joint Planck+WMAP9+BAO data
\cite{Ade13a,Ade13b}. By this comparison we find some constraint on
the Gauss-Bonnet coupling parameter $\alpha_{_{GB}}$.

\section{The Setup}

The 4-dimensional action for a DBI model in the presence of the
Gauss-Bonnet term, which is non-minimally coupled to the DBI field,
can be written as follows
\begin{equation}
S=\int\sqrt{-g}\Bigg[\,\frac{1}{\kappa^{2}}R-f^{-1}(\phi)\sqrt{1-f(\phi)\partial_{\alpha}\phi\partial^{\alpha}\phi}
-V(\phi)+\alpha(\phi){\cal{L}}_{GB}\,\Bigg]d^{4}x\,, \label{1}
\end{equation}
where $R$ is the 4-dimensional Ricci scalar, $\phi$ is the DBI field
and $V(\phi)$ is its potential. $f^{-1}(\phi)$, which is  the
inverse brane tension, is related to the geometry of the throat.
$\alpha(\phi)$ is a potential term which is coupled to the
Gauss-Bonnet term. Also, ${\cal{L}}_{GB}$, the lagrangian term
corresponding to the Gauss-Bonnet effect, is given by
\begin{equation}
\label{2}
{\cal{L}}_{GB}=R^{2}-4R_{\mu\nu}R^{\mu\nu}+R_{\mu\nu\alpha\beta}R^{\mu\nu\alpha\beta}\,.
\end{equation}

Einstein's field equations obtained from action~(\ref{1}) are given
by the following expression
\begin{equation}
\label{3}
G_{\mu\nu}=\kappa^{2}T_{\mu\nu}+\kappa^{2}{\cal{T}}_{\mu\nu}\,,
\end{equation}
where $T_{\mu\nu}$, the energy momentum tensor corresponding to the
DBI field is given by
\begin{equation}
\label{4} T_{\mu\nu}=-\frac{\partial_{\mu}\phi
\partial_{\nu}\phi}{\sqrt{1-f(\phi)\partial_{\alpha}\phi\partial^{\alpha}\phi}}+g_{\mu\nu}\Big(-f^{-1}(\phi)
\sqrt{1-f(\phi)\partial_{\alpha}\phi\partial^{\alpha}\phi}+f^{-1}(\phi)-V(\phi)\Big)\,,
\end{equation}
and ${\cal{T}}_{\mu\nu}$ is the energy momentum tensor corresponding
to the Gauss-Bonnet term, given by
\begin{eqnarray}
{\cal{T}}^{\mu\nu}=\frac{1}{2}g^{\mu\nu}\alpha(\phi){\cal{L}}_{GB}-2\alpha(\phi)RR^{\mu\nu}+4\alpha(\phi)R^{\mu}_{\rho}R^{\nu\rho}
-2\alpha(\phi)R^{\mu\rho\sigma\tau}R^{\nu}_{\rho\sigma\tau}-4\alpha(\phi)R^{\mu\rho\sigma\nu}R_{\rho\sigma}\nonumber
\end{eqnarray}
\begin{eqnarray}
+2\Big[\nonumber\nabla^{\mu}\nabla^{\nu}\,\alpha(\phi)\Big]R-2g^{\mu\nu}\Big[\nabla^{2}\alpha(\phi)\Big]R
-4\Big[\nabla_{\rho}\nabla^{\mu}\alpha(\phi)\Big]R^{\nu\rho}-4\Big[\nabla_{\rho}\nabla^{\nu}\alpha(\phi)\Big]R^{\mu\rho}
+4\Big[\nabla^{2}\alpha(\phi)\Big]R^{\mu\nu}
\end{eqnarray}
\begin{equation}
+4g^{\mu\nu}\Big[\nabla_{\rho}\nabla_{\sigma}\alpha(\phi)\Big]R^{\rho\sigma}
-4\Big[\nabla_{\rho}\nabla_{\sigma}\alpha(\phi)\Big]R^{\mu\rho\sigma\nu}\,.
\label{5}
\end{equation}

The total energy momentum tensor ($T_{\mu\nu}+{\cal{T}}_{\mu\nu}$)
leads to the following energy density and pressure for this model
\begin{equation}
\label{6}
\rho=\frac{-f^{-1}\sqrt{1-f\dot{\phi}^{2}}+f^{-1}}{\sqrt{1-f\dot{\phi}^{2}}}+V-12H^{3}\alpha'\dot{\phi}\,,
\end{equation}
\begin{equation}
\label{7}
p=-f^{-1}\,\sqrt{1-f\dot{\phi}^{2}}\,+f^{-1}-V-12H^{2}\Big(\alpha''\dot{\phi}^{2}+\alpha'\ddot{\phi}\Big)
-24H\dot{H}\alpha'\dot{\phi}-24H^{3}\alpha'\dot{\phi}\,,
\end{equation}
where, a dot shows derivative with respect to the time and a prime
marks derivative with respect to the DBI field. By assuming the
following spatially flat FRW line element
\begin{equation}
\label{8} ds^{2}=-n^{2}(t)dt^{2}+a^{2}(t)\gamma_{ij}dx^{i}dx^{j}\,,
\end{equation}
where $\gamma_{ij}$ is a maximally symmetric 3-dimensional metric
defined as $\gamma_{ij}=\delta_{ij}+k\frac{x_{i}x_{j}}{1-kr^{2}}$
where $k=-1,0,+1$ parameterizes the spatial curvature and
$r^{2}=x_{i}x^{i}$. We assume $n^{2}(t)=1$ and consider the $(0,0)$
component of the Einstein's field equations in order to obtain the
Friedmann equation for this model as
\begin{equation}
\label{9}
H^{2}=\frac{\kappa^{2}}{3}\,\Bigg[\frac{-f^{-1}\sqrt{1-f\dot{\phi}^{2}}+f^{-1}}{\sqrt{1-f\dot{\phi}^{2}}}+V
+12H^{3} \alpha'\dot{\phi}\Bigg]\,.
\end{equation}
By varying the action (\ref{1}) with respect to the scalar field we
reach the following equation of motion
\begin{equation}
\label{10}
\frac{\ddot{\phi}}{(1-f\dot{\phi}^{2})^{\frac{3}{2}}}+\frac{3H\dot{\phi}}{(1-f\dot{\phi}^{2})^{\frac{1}{2}}}-V'
=-f'f^{-2}\Bigg[1+\sqrt{1-f\dot{\phi}^{2}}-\frac{1}{2}\frac{f\,\dot{\phi}^{2}}{\sqrt{1-f\dot{\phi}^{2}}}\Bigg]+12\alpha'
H^{2}\Big(\dot{H}+H^{2}\Big)\,.
\end{equation}

If we consider the slow-roll approximation, where $\dot{\phi}^{2}\ll
1$ and $\ddot{\phi}\ll|3H\dot{\phi}|$, the energy density and the
equation of motion of the DBI field take the following forms
respectively
\begin{equation}
\label{11} \rho=V-12H^{3}\alpha'\dot{\phi}\,,
\end{equation}
and
\begin{equation}
\label{12} 3H\dot{\phi}-V'+2f'f^{-2}-\alpha'R_{GB}=0\,,
\end{equation}
where
\begin{equation}
\label{13} R_{GB}=12H^{2}\Big(\dot{H}+H^{2}\Big)\,.
\end{equation}
In this regard, the Friedmann equation of the model (\ref{9})
reduces to the following expression
\begin{equation}
\label{14}
H^{2}=\frac{\kappa^{2}}{3}\Bigg[V-12H^{3}\alpha'\dot{\phi}\Bigg]\,.
\end{equation}
The slow-roll parameters which are defined by $\epsilon\equiv
-\frac{\dot{H}}{H^{2}}$ and $\eta\equiv
-\frac{1}{H}\frac{\ddot{H}}{\dot{H}}$, in this model take the
following forms respectively
\begin{equation}
\label{15}
\epsilon=\frac{1}{2\kappa^{2}}\frac{V'^{2}}{V^{2}}\frac{\frac{2f'f^{-2}}{V'}-1-{\alpha'R_{GB}}{V'}}
{\Big(1-\frac{4\kappa^{2}}{3}\alpha'\big(V'-2f'f^{-2}+\alpha'R_{GB}\big)\Big)^{2}}
\Bigg[\frac{1}{1-\frac{4\kappa^{2}}{3}\alpha'\big(V'-2f'f^{-2}+\alpha'R_{GB}\big)}-\frac{4V}{3V'}X\Bigg]\,,
\end{equation}
where
\begin{equation}
\label{16}
X=\frac{\alpha''\big(V'-2f'f^{-2}+\alpha'R_{GB}\big)+\alpha'\big(V''-2f''f^{-2}+4f'^{2}f^{-3}+\alpha''R_{GB}\big)}
{\Big(1-\frac{4\kappa^{2}}{3}\alpha'\big(V'-2f'f^{-2}+\alpha'R_{GB}\big)\Big)^{2}}\,,
\end{equation}
and
\begin{equation}
\label{17}
\eta=-\frac{2\dot{Y}}{HY}-\frac{1}{3H^{2}\alpha'}\frac{1-Z}{Z}\Bigg[\alpha''\big(V'-2f'f^{-2}+\alpha'R_{GB}\big)+
\alpha'\big(V''-2f''f^{-2}+4f'^{2}f^{-3}+\alpha''R_{GB}\big)\Bigg]\,,
\end{equation}
where
\begin{eqnarray}
Y=\Bigg[V'-2f'f^{-2}+\alpha'R_{GB}\Bigg]\times\nonumber
\end{eqnarray}
\begin{equation}
\Bigg[\frac{\kappa^{2}V'}{9H}+\frac{4\kappa^{2}}{9}\alpha''
\big(V'-2f'f^{-2}+\alpha'R_{GB}\big)+\frac{4\kappa^{2}}{9}\alpha'
\Big(V''-2f''f^{-2}+4f'^{2}f^{-3}+\alpha''R_{GB}\Big)\Bigg]\,,
\label{18}
\end{equation}
and
\begin{equation}
Z=1-\frac{4\kappa^{2}}{3}\,\alpha'\,\Big(V'-2f'f^{-2}+\alpha'
R_{_GB}\Big) \label{18-2}
\end{equation}

The inflation takes place under the condition for which
$\{\epsilon,\eta\}<1$; as soon as one of these slow-roll parameters
reaches the unity, the inflationary phase terminates.

The number of e-folds during inflation is defined as
\begin{equation}
\label{19} N=\int_{t_{hc}}^{t_{f}} H dt\,,
\end{equation}
which, in our setup and within the slow-roll approximation can be
expressed as
\begin{equation}
\label{20} N\simeq \int_{\phi_{hc}}^{\phi_{f}}
\frac{3H^{2}}{V'-2f'f^{-2}+\alpha'R_{GB}} d\phi\,,
\end{equation}
where $\phi_{hc}$ denotes the value of the field when the universe
scale observed today crosses the Hubble horizon during inflation and
$\phi_{f}$ is the value of the field when the universe exits the
inflationary phase. In a model with a DBI field which is
non-minimally coupled to the Gauss-Bonnet term, the number of
e-folds in the slow-roll approximation takes the following form
\begin{equation}
\label{21} N\simeq-\int_{\phi_{hc}}^{\phi_{f}}
\frac{\kappa^{2}V\Big[1-\frac{4\kappa^{2}}{3}\alpha'\big(V'-2f'f^{-2}+\alpha'R_{GB}\big)\Big]^{-1}}{V'-2f'f^{-2}+\alpha'R_{GB}}
d\phi\,.
\end{equation}
In the next section, we study the scalar perturbation of the metric
(the density perturbation) with details.

\section{Perturbations}

In this section, we study the linear perturbation theory in the
presence of the DBI field and the Gauss-Bonnet term in the action.
Among many different ways, depending on the choice of gauge
(coordinates) characterizing the cosmological perturbations, we
choose the longitudinal gauge in which the scalar metric
perturbations of the FRW background are given
by~\cite{Bar80,Muk92,Ber95}
\begin{equation}
\label{22}
ds^{2}=-\big(1+2\Phi\big)dt^{2}+a^{2}(t)\big(1-2\Psi\big)\delta_{i\,j}\,dx^{i}dx^{j}\,,
\end{equation}
where $a(t)$ is the scale factor, $\Phi=\Phi(t,x)$ and
$\Psi=\Psi(t,x)$, the metric perturbations, are gauge-invariant
variables. The form of the spatial dependence of all perturbed
quantities is similar to the plane waves $e^{ikx}$, where $k$ is the
wave number. Any perturbation of the metric, through Einstein's
field equations, leads to the perturbation in the energy-momentum
tensor. For the perturbed metric (\ref{22}), the perturbed
Einstein's field equations can be obtained as follows
\begin{equation}
\label{23}
-6H(H\Phi+\dot{\Psi})-\frac{2k^{2}}{a^{2}}=\kappa^{2}\frac{f'\delta\phi}{f^{2}}\Big(1-\frac{1}{\sqrt{1-\dot{\phi}^{2}}}\Big)+
\kappa^{2}\frac{f'\dot{\phi}^{2}\delta\phi-2f\dot{\phi}\dot{\delta\phi}-2f\dot{\phi}^{2}\Phi}{2f(1-f\dot{\phi}^{2})^{\frac{3}{2}}}
+\kappa^{2}V'\delta\phi+\kappa^{2}\delta \rho_{_{GB}}\,,
\end{equation}
\begin{eqnarray}
\label{24}
2\ddot{\Psi}+6H(H\Phi+\dot{\Psi})+2H\dot{\Phi}+4\dot{H}\Phi+\frac{2}{3a^{2}}k^{2}(\Phi-\Psi)=\nonumber
\end{eqnarray}
\begin{equation}
\kappa^{2}\frac{f'\delta\phi}{f^{2}}\Big(\sqrt{1-f\dot{\phi}^{2}}-1\Big)+\kappa^{2}\frac{f'\dot{\phi}^{2}\delta\phi
-2f\dot{\phi}\dot{\delta\phi}-2f\dot{\phi}^{2}\Phi}{2f\sqrt{1-f\dot{\phi}^{2}}}-\kappa^{2}V'\delta\phi+\kappa^{2}\delta
p_{_{GB}}\,,
\end{equation}
\begin{equation}
\label{25}
\dot{\Psi}+H\Phi=-\frac{\kappa^{2}V(\phi)}{\sqrt{1-\dot{\phi}^{2}}}\frac{\dot{\phi}\delta\phi}{2}
+4H^{2}\big(\alpha''\dot{\phi}+\alpha'\ddot{\phi}\big)\delta\phi-4H^{3}\alpha'\delta\phi
-\frac{8H}{3}\alpha'\dot{\phi}\big(3H\Phi-3\dot{\Psi}\big)-4H^{2}\alpha'\dot{\phi}\Phi\,,
\end{equation}
\begin{equation}
\label{26}
\Psi-\Phi=-4\big(\alpha''\dot{\phi}^{2}+\alpha'\ddot{\phi}\big)\Psi-4H\alpha'\dot{\phi}\Phi
-4\big(\dot{H}+H^{2}\big)\alpha'\delta\phi\,.
\end{equation}
As we see from equation (\ref{26}), in the model with non-minimal
coupling between the DBI field and the Gauss-Bonnet term, the two
metric perturbations are not equal. In equations (\ref{23}) and
(\ref{24}), the perturbed energy density and pressure of the
Gauss-Bonnet term are given by the following expression
\begin{equation}
\label{27}
\delta\rho_{_{GB}}=-12H^{3}\big(\alpha''\dot{\phi}+\alpha'\ddot{\phi}\big)\delta\phi
+12H^{2}\alpha'\dot{\phi}\big(4H\Phi-3\dot{\Psi}\big)+4H\frac{k^{2}}{a^{2}}\big(H\alpha'\delta\phi+2\alpha'\dot{\phi}\Psi\big)\,,
\end{equation}
and
\begin{eqnarray}
\delta
p_{_{GB}}=-12H^{2}\ddot{\delta\alpha}-24H\Big(\dot{H}+H^{2}\Big)\alpha'\delta\phi
-\frac{8k^{2}}{a^{2}}\Big(\dot{H}+H^{2}\Big)\alpha'\delta\phi+8H\alpha'\dot{\phi}\Big(3\dot{H}\Phi+3H\dot{\Phi}-3\ddot{\Psi}\Big)
\nonumber
\end{eqnarray}
\begin{eqnarray}
+8\Big(H(\alpha''\dot{\phi}^{2}+\alpha'\ddot{\phi})+\dot{H}\alpha'\dot{\phi}+3H^{2}\alpha'\dot{\phi}\Big)
+24H\Big(H(\alpha''\dot{\phi}^{2}+\alpha'\ddot{\phi})
+2\dot{H}+\alpha'\dot{\phi}+H^{2}\alpha'\dot{\phi}\Big)\Phi\nonumber
\end{eqnarray}
\begin{equation}
-\frac{8k^{2}}{a^{2}}\Big((\alpha''\dot{\phi}^{2}+\alpha'\ddot{\phi})\Psi+H\alpha'\dot{\phi}\,\Phi\Big)
+12H^{2}\alpha'\dot{\phi}\,\dot{\Phi}\,, \label{28}
\end{equation}
which are obtained by perturbing the (0,0) and (i,j) components of
equation (\ref{5}). By varying the scalar field's equation of motion
(\ref{10}) we achieve
\begin{eqnarray}
\ddot{\delta\phi}+3H\dot{\delta\phi}+2V'\Phi-\dot{\phi}\Big(\dot{\Phi}+3\dot{\Psi}\Big)+f''f^{-2}
\Big(1-\frac{1}{2}f\dot{\phi}^{2}\Big)\delta\phi-2f'^{2}f^{-3}\Big(1-f\dot{\phi}^{2}\Big)\delta\phi
-V''\sqrt{1-f\dot{\phi}^{2}}\nonumber
\end{eqnarray}
\begin{eqnarray}
+f'f^{-1}\Big(\dot{\phi}\dot{\delta\phi}+\dot{\phi}^{2}\Phi\Big)
-\frac{f'^{2}f^{2}\dot{\phi}^{2}\delta\phi-2f'f^{-1}\big(\dot{\phi}\dot{\delta\phi}+\dot{\phi}^{2}\Phi\big)
-V'f'\dot{\phi}^{2}\delta\phi-2V'f\big(\dot{\phi}\dot{\delta\phi}-\dot{\phi}^{2}\Phi\big)}
{2\,\sqrt{1-f\dot{\phi}^{2}}}\nonumber
\end{eqnarray}
\begin{eqnarray}
-6\alpha'H^{2}\Big(\dot{H}+H^{2}\Big)\Bigg[f'\dot{\phi}^{2}\delta\phi
-2f\big(\dot{\phi}\dot{\delta\phi}-\dot{\phi}^{2}\Phi\big)\Bigg]+\sqrt{1-f\dot{\phi}^{2}}
\Bigg[2H^{2}\delta
R-8\dot{H}\Big(H\big(3H\Phi-3\dot{\Psi}\big)-\frac{k^{2}}{a^{2}}\Psi
\Big)\Bigg]\nonumber
\end{eqnarray}
\begin{equation}
=\frac{f'\delta\phi\ddot{\phi}\dot{\phi}^{2}-2f\ddot{\phi}\dot{\phi}\dot{\delta\phi}-
2f^{2}\dot{\phi}^{3}\ddot{\phi}\Phi}{\big(1-f\dot{\phi}^{2}\big)^{2}}\,,
\label{29}
\end{equation}

where the variation of the Ricci scalar is defined as
\begin{equation}
\label{30} \delta
R=2\Bigg[2\frac{k^{2}}{a^{2}}\Psi-\Big(3\dot{H}\Phi+3H\dot{\Phi}-3\ddot{\Psi}\Big)
-4H\Big(3H\Phi-3\dot{\Psi}\Big)-\Big(3\dot{H}-\frac{k^{2}}{a^{2}}\Big)\Phi\Bigg]\,.
\end{equation}

One can decompose the scalar perturbations into two parts. One part
which is the projection parallel to the trajectory is called
adiabatic or curvature perturbations (if there is only one scalar
field during the inflationary period, we deal with this type of
perturbations ~\cite{Bas99,Gor01,Lop04,Kal05,Maa00}). Another part
which is the projection orthogonal to the trajectory is dubbed the
entropy or isocurvature perturbations (if inflation is driven by
more than one scalar field ~\cite{Bas99,Gor01,Lan00,Lan07} or it
interacts with other fields such as the scalar Ricci term
~\cite{Lop04,Kal05}, we deal with this type of perturbations). In
this work, since the DBI field is non-minimally coupled to the
Gauss-Bonnet term, the perturbations are expected to be
non-adiabatic. To explore the first order cosmological perturbation
(linear perturbation), we can define a gauge-invariant primordial
curvature perturbation $\zeta$, on scales outside the horizon, as
follows ~\cite{Bar83}
\begin{equation}
\label{31} \zeta=\Psi-\frac{H}{\dot{\rho}}\delta\rho\,\,.
\end{equation}
The above quantity, on uniform density hypersurfaces where
$\delta\rho=0$, reduces to the curvature perturbation, $\Psi$. By
using the equation (\ref{31}) one can obtain the following equation
for time evolution of $\zeta$ \cite{Wan00}
\begin{equation}
\label{32} \dot{\zeta}=H\left(\frac{\delta
p_{nad}}{\rho+p}\right)\,.
\end{equation}
Equation (\ref{32}) shows that, independent of the form of the
gravitational field equations, any change in the curvature
perturbation on uniform-density hypersurfaces, on large scales, is
due to the non-adiabatic part of the pressure perturbation. $\zeta$
is constant if the pressure perturbation is adiabatic on the large
scales. In our setup, since the non-adiabatic perturbation is
expected, the curvature perturbation would vary with time.

In general, the pressure perturbation (in any gauge) can be
decomposed into adiabatic and entropic (non-adiabatic) parts
~\cite{Wan00}
\begin{equation}
\label{33} \delta p=c_{s}^{2}\,\delta\rho+\dot{p}\,\Gamma\,\,,
\end{equation}
where $c_{s}^{2}=\frac{\dot{p}}{\dot{\rho}}$ is the sound effective
velocity. In equation (\ref{33}), $\delta
p_{nad}=\dot{p}\,\Gamma$\,, is the non-adiabatic part, where
$\Gamma$ marks the displacement between hypersurfaces of uniform
pressure and density.

In the presence of the non-minimal coupling between the DBI field
and the Gauss-Bonnet term, $\delta p_{nad}$ is not zero anymore. So,
from equation (\ref{33}), we can find the $\delta p_{nad}$ as
follows
\begin{eqnarray}
\delta p_{nad}=\kappa^{2}V'\Big(f\dot{\phi}^{2}-2\Big)\delta\phi+
\frac{2}{\kappa^{2}}\Bigg[1-f\dot{\phi}^{2}-D\Bigg]\Bigg[-3H\Big(H\Phi+\dot{\Psi}\Big)-\frac{k^{2}}{a^{2}}\Bigg]\nonumber
\end{eqnarray}
\begin{equation}
-\delta\rho_{_{GB}}\Big(1-f\dot{\phi}^{2}\Big) -2\kappa^{2}\frac{f'
\delta
\phi}{f^{2}}\Big(1-f\dot{\phi}^{2}-\sqrt{1-f\dot{\phi}^{2}}\Big)+\delta
p_{_{GB}}\,,
\label{34}
\end{equation}
where
\begin{equation}
D=\frac{f'f^{-2}\dot{\phi}\sqrt{1-f\dot{\phi}^{2}}+\frac{2\dot{\phi}\ddot{\phi}+f'f^{-1}\dot{\phi}^{3}}
{\sqrt{1-f\dot{\phi}^{2}}}-f'f^{-2}\dot{\phi}-V'\dot{\phi}-\dot{p}_{_{GB}}}
{\frac{f'\dot{\phi}}{f^{2}}-\frac{f'\dot{\phi}}{f^{2}\sqrt{1-f\dot{\phi}^{2}}}+
\frac{f'\dot{\phi}^{3}-2f\dot{\phi}\ddot{\phi}}{2f(1-f\dot{\phi}^{2})^{\frac{3}{2}}}+V'\dot{\phi}+\dot{\rho}_{_{GB}}}\,.
\label{35}
\end{equation}
From equation (\ref{32}), we find that, this non-vanishing,
non-adiabatic pressure leads to the non-vanishing time evolution of
the primordial curvature perturbation as follows
\begin{eqnarray}
\dot{\zeta}=\frac{Hf\dot{\phi}^{2}}{f\dot{\phi}^{2}+(\rho_{_{GB}}+p_{_{GB}})f\sqrt{1-f\dot{\phi}^{2}}}
\Bigg\{\frac{2}{\kappa^{2}}\Bigg[1-f\dot{\phi}^{2}-D\Bigg]\Bigg[-3H\Big(H\Phi+\dot{\Psi}\Big)-\frac{k^{2}}{a^{2}}\Bigg]
\nonumber
\end{eqnarray}
\begin{equation}
+\kappa^{2}V'\Big(f\dot{\phi}^{2}-2\Big)\delta\phi
-\delta\rho_{_{GB}}\Big(1-f\dot{\phi}^{2}\Big) -2\kappa^{2}\frac{f'
\delta
\phi}{f^{2}}\Big(1-f\dot{\phi}^{2}-\sqrt{1-f\dot{\phi}^{2}}\Big)+\delta
p_{_{GB}}\Bigg\}\,.
\label{36}
\end{equation}

We see that, in the presence of the non-minimally coupled
Gauss-Bonnet term, the primordial curvature perturbations attain an
explicit time-dependence.

The scales of cosmological interest have spent most of their time
far outside the Hubble radius and have re-entered only relatively
recently in the Universe history. So, in order to obtain scalar and
tensorial perturbations in our model, it is enough to consider the
slow-roll approximation at the large scales, $k\ll aH$.  In this
scale, $\ddot{\Phi}$, $\ddot{\Psi}$, $\dot{\Phi}$ and $\dot{\Psi}$
are negligible (see ~\cite{Amn06,Amn07,Noz12,Noz13}). So, at large
scales, the perturbed equation of motion takes the following form
\begin{eqnarray}
\label{37}
3H\delta\dot{\phi}+\Bigg[f''f^{-2}-2f'^{2}f^{-3}\delta\phi-V''\Bigg]\delta\phi+12\alpha''R_{GB}\delta\phi
=-2V'\Phi+2H^{2}\delta R-24\dot{H}H^{2}\Phi\,.
\end{eqnarray}
Also, the perturbed Einstein's field equation (\ref{25}) gives
\begin{equation}
\label{38}
\Phi=\frac{\big(4H^{2}\alpha''-\kappa^{2}\big)\dot{\phi}\delta\phi-4H^{3}\alpha'\delta\phi}{2H+12H^{2}\alpha'\dot{\phi}}\,.
\end{equation}
So, we can  rewrite the equation (\ref{37}) as follows
\begin{eqnarray}
3H\delta\dot{\phi}=\Bigg[V''-f''f^{-2}+2f'^{2}f^{-3}\Bigg]\delta\phi\nonumber
\end{eqnarray}
\begin{equation}
+\Bigg[12H^{2}\Big(4H^{2}-\dot{H}\Big)+24H\dot{H}-2V'\Bigg]\frac{\big(4H^{2}\alpha''-\kappa^{2}\big)\dot{\phi}\delta\phi
-4H^{3}\alpha'\delta\phi}{2H+12H^{2}\alpha'\dot{\phi}}\,.
\label{39}
\end{equation}

In order to solve the equation (\ref{39}), to obtain the explicit
form for the perturbed field $\delta\phi$, we introduce the function
${\cal{A}}$ as
\begin{equation}
\label{40} {\cal{A}}\equiv\frac{V\delta\varphi}{V'}\,,
\end{equation}
by which we rewrite equation (\ref{39}) as follows
\begin{equation}
\label{41} \frac{{\cal{A}}'}{{\cal{A}}}=\frac{V'}{V}-\frac{V''}{V}
+\frac{12\alpha''R_{GB}+V''-f''f^{-2}+2f'^{2}f^{-3}}{\frac{1}{2}\alpha'R_{GB}-2f'f^{-2}+V'}+{\cal{D}}
\end{equation}
where
\begin{eqnarray}
{\cal{D}}=\Bigg[12H^{2}\Big(4H^{2}+\epsilon H^{2}\Big)-24\epsilon
H^{3}-2V'\Bigg]\Bigg[\frac{4H^{2}\alpha''-\kappa^{2}}{6H^{2}+12H^{2}\alpha'\Big(\frac{1}{2}\alpha'R_{GB}-2f'f^{-2}+V'\Big)}
\nonumber
\end{eqnarray}
\begin{equation}
-\frac{4H^{3}\alpha'}{2H\Big(\frac{1}{2}\alpha'R_{GB}-2f'f^{-2}+V'\Big)+
4H\alpha'\Big(\frac{1}{2}\alpha'R_{GB}-2f'f^{-2}+V'\Big)}\Bigg]\,.
\label{42}
\end{equation}
A solution of this equation is given by the following expression
\begin{equation}
\label{43}
{\cal{A}}={\cal{C}}\exp\bigg(\int\frac{{\cal{A}}'}{{\cal{A}}}d\varphi\bigg)\,,
\end{equation}
where ${\cal{C}}$ is an integration constant. So, from equation
(\ref{40}) we achieve
\begin{equation}
\label{44} \delta\phi=\frac{{\cal{C}}V'}{V}\exp
\Bigg[\int\Bigg(\frac{V'}{V}-\frac{V''}{V}
+\frac{12\alpha''R_{GB}+V''-f''f^{-2}+2f'^{2}f^{-3}}{\frac{1}{2}\alpha'R_{GB}-2f'f^{-2}+V'}+{\cal{D}}\Bigg)d\phi\Bigg]\,.
\end{equation}
By using equation (\ref{44}), we can find the following expression
for the density perturbation amplitude
\begin{equation} \label{45}
A_{s}^{2}=\frac{k^{3}{\cal{C}}}{2\pi^{2}}\frac{V'^{2}}{V^{2}} \exp
\Bigg[2\int\Bigg(\frac{V'}{V}-\frac{V''}{V}
+\frac{12\alpha''R_{GB}+V''-f''f^{-2}+2f'^{2}f^{-3}}{\frac{1}{2}\alpha'R_{GB}-2f'f^{-2}+V'}+{\cal{D}}\Bigg)d\phi\Bigg]\,.
\end{equation}

The scale-dependence of the perturbations is described by the
spectral index as
\begin{equation}
\label{46} n_{s}-1=\frac{d \ln A_{s}^{2}}{d \ln k}\,.
\end{equation}
where the interval in wave number is related to the number of
e-folds by the relation $ d \ln k(\phi)=d N(\phi)$. By using
equations (\ref{43}) and (\ref{46}), the scalar spectral index
becomes
\begin{eqnarray}
n_{s}-1=\frac{1}{3H^{2}}\Bigg[f''f^{-2}-12\alpha''R_{GB}-V''-2f'^{2}f^{-3}\Bigg]
+\Bigg[24\epsilon H^{3}+2V'-12H^{2}\Big(4H^{2}+\epsilon
H^{2}\Big)\Bigg]\nonumber
\end{eqnarray}
\begin{equation}
\times
\frac{\Big(4H^{2}\alpha''-\kappa^{2}\Big)\Big(\frac{1}{2}\alpha'R_{GB}-2f'f^{-2}+V'\Big)-12H^{4}\alpha'}
{9H^{2}\Big(2H+H\alpha'\big(\frac{1}{2}\alpha'R_{GB}-2f'f^{-2}+V'\big)\Big)}
\label{47}
\end{equation}

Also, the tensor perturbations amplitude of a given mode when
leaving the Hubble radius is defined as
\begin{equation}
\label{48} A_{T}^{2}=\frac{4\kappa^{2}}{25\pi}H^{2}\Bigg|_{k=aH}\,.
\end{equation}
In a model with a DBI field which is non-minimally coupled to the
Gauss-Bonnet term, the tensor perturbations amplitude of a given
mode when leaving the Hubble radius is given by
\begin{equation}
\label{49} A_{T}^{2}=\frac{4\kappa^4
V}{75\pi\big(1-\kappa^{2}f\big)}\Bigg[1-\frac{4\kappa^{2}}{3}\alpha'\Big(V'-2f'f^{-2}+\alpha'R_{GB}\Big)\Bigg]^{-1}\,.
\end{equation}
So, the tensor spectral index which is defined as
\begin{equation}
\label{50} n_{T}=\frac{d \ln A_{T}^{2}}{d \ln k}\,,
\end{equation}
in this setup takes the following form
\begin{eqnarray}
n_{T}=\frac{\kappa^{2}\dot{\phi}}{3H^{2}}\Bigg\{V'\Bigg[\frac{4\kappa^{2}}{3}\alpha'\Big(V'
-2f'f^{-2}+\alpha'R_{GB}\Big)\Bigg]+V\Bigg[1-\frac{4\kappa^{2}}{3}\alpha'\Big(V'
-2f'f^{-2}+\alpha'R_{GB}\Big)\Bigg]^{-2} \nonumber
\end{eqnarray}
\begin{eqnarray}
\times \Bigg[\frac{4\kappa^{2}}{3}\alpha''\Big(V'
-2f'f^{-2}+\alpha'R_{GB}\Big)+\frac{4\kappa^{2}}{3}\alpha'\Big(V''-2f''f^{-2}+4f'^{2}f^{-3}+\alpha''R_{GB}\Big)\Bigg]\Bigg\}\,.
\label{51}
\end{eqnarray}

The ratio between the amplitudes of tensor and scalar perturbations
(tensor-to-scalar ratio) is another important parameter which is
given by
\begin{equation}
\label{44} r=\frac{A_{T}^{2}}{A_{s}^{2}}=
\frac{8\kappa^{4}\pi}{75k^{3}{\cal{C}}}\frac{V^{3}}{V'^{2}}
\times\frac{ \exp \Bigg[-2\int\Bigg(\frac{V'}{V}-\frac{V''}{V}
+\frac{12\alpha''R_{GB}+V''-f''f^{-2}+2f'^{2}f^{-3}}{\frac{1}{2}\alpha'R_{GB}-2f'f^{-2}+V'}+{\cal{D}}\Bigg)d\phi\Bigg]}
{1-\frac{4\kappa^{2}}{3}\alpha'\Big(V'-2f'f^{-2}+\alpha'R_{GB}\Big)}\,.\label{52}
\end{equation}

\section{Non-Gaussianity}

In this section we are going to study the non-Gaussianity of the
primordial density perturbation in this model. As we have said in
the Introduction, for a Gaussian distribution, the three-point
function and also other odd correlation functions are zero. But this
is not the case for non-Gaussian distribution. So, to study the
primordial non-Gaussianitiy, we study the three-point correlators.
To this end, we should expand the action (\ref{1}) up to the cubic
order in the small fluctuations (which have their origin in the
quantum behavior of both the field $\phi$ and the space-time metric,
$g_{\mu\nu}$) around the homogeneous background solution. These
cubic terms in lagrangian, lead to a change both in the ground state
of the quantum field and non-linearities in the evolution
\cite{Mal03}. To compute the Einstein action to the third order, we
work in the ADM metric formalism \cite{Arn60}
\begin{equation}
\label{53}
ds^{2}=-N^{2}dt^{2}+h_{ij}\big(dx^{i}+N^{i}dt\big)\big(dx^{j}+N^{j}dt\big)
\end{equation}
where $N$ and $N^{i}$ are the lapse and shift functions,
respectively. It should be noticed that we consider only scalar
metric perturbations about the flat FRW background. A general
parametrization of the scalar fluctuations in the metric is provided
by expanding the lapse function $N$ and the shift vector $N^{i}$, as
$N=1+2\Phi$ and $N^{i}=\delta^{ij}\partial_{j}B$. Note that there is
no need to know $N$ or $N^{i}$ up to the second order. The reason is
that the second order equation is multiplied by a factor which
vanishes by the first order solution. Also, the contribution of the
third order term vanishes because it is multiplied by the constraint
equation at the zeroth order (the zeroth order solution obeys the
equations of motion) \cite{Mal03,Fel11a,Koy10}. In which follows, we
work in the uniform-field gauge for which $\delta\phi=0$. This gauge
fixes the time-component of a gauge-transformation vector
$\xi^{\mu}$ and so $h_{ij}$ can be written as
$a^{2}(t)\,e^{2\Psi}\delta^{ij}$ \cite{Mal03,Lyt05}. Now, we write
the perturbed metric at the linear level, as
\cite{Bar80,Muk92,Ber95}
\begin{equation}
\label{54}
ds^{2}=-\big(1+2\Phi\big)dt^{2}+2a^{2}(t)B_{,i}dx^{i}dt+a^{2}(t)\big(1-2\Psi\big)\delta_{ij}dx^{i}dx^{j}
\end{equation}
By expanding the action (\ref{1}) up to the second order, we obtain
\begin{eqnarray}
S_{2}=\int dt\, d^{3}x\,
a^{3}\Bigg[\bigg(24H\alpha'\dot{\phi}-\frac{3}{\kappa^{2}}\bigg)\dot{\Psi}^{2}
-\frac{\bigg(2H\alpha'\dot{\phi}-\frac{2}{\kappa^{2}}\bigg)}{a^{2}}\dot{\Psi}\partial^{2}B
-\frac{\bigg(\frac{2H}{\kappa^{2}}
-24H^{2}\alpha'\dot{\phi}\bigg)}{a^{2}}\Phi\,\partial^{2}B \nonumber
\end{eqnarray}
\begin{eqnarray}
-\frac{2}{a^{2}}\bigg(\frac{1}{\kappa^{2}}
-8H\alpha'\dot{\phi}\bigg)\Phi\,\partial^{2}\Psi
+\bigg(\frac{6H}{\kappa^{2}}
-72H^{2}\alpha'\dot{\phi}\bigg)\Phi\dot{\Psi}+\frac{1}{a^{2}}
\bigg(\frac{1}{\kappa^{2}}-8\Big(\alpha''\dot{\phi}^{2}+\alpha'\ddot{\phi}\Big)\partial_{i}\Psi\partial
^{i}\Psi\bigg)\times\nonumber
\end{eqnarray}
\begin{eqnarray}
\bigg(48H^{3}\alpha'\dot{\phi}-\frac{3H^{2}}{\kappa^{2}}+\frac{\dot{\phi}^{2}}{2\sqrt{1-f\dot{\phi}^{2}}}+
\frac{f\,\dot{\phi}^{4}}{2\big(1-f\dot{\phi}^{2}\big)^{\frac{3}{2}}}\bigg)\Phi^{2}\Bigg]\,.
\label{55}
\end{eqnarray}

By using the above second order equation, we can find the equation
of motion of $\Phi$ and $B$ respectively as follows
\begin{equation}
\label{56}
\Phi=\frac{2\kappa^{-2}-16H\dot{\alpha}}{2\kappa^{-2}H-24H^{2}\alpha'\dot{\phi}}\,\dot{\Psi}\,,
\end{equation}
\begin{equation}
\label{57}
\frac{1}{a^{2}}\partial^{2}B=\frac{2\bigg(48H^{3}\alpha'\dot{\phi}
-\frac{3H^{2}}{\kappa^{2}}+\frac{\dot{\phi}^{2}}{2\sqrt{1-f\dot{\phi}^{2}}}+
\frac{f\,\dot{\phi}^{4}}{2\big(1-f\dot{\phi}^{2}\big)^{\frac{3}{2}}}\bigg)}{3\bigg(\frac{2H}{\kappa^{2}}
-24H^{2}\alpha'\dot{\phi}\bigg)}\Phi+3\dot{\Psi}-\frac{2\bigg(24H\alpha'\dot{\phi}-\frac{3}{\kappa^{2}}\bigg)}
{a^{2}\bigg(\frac{2H}{\kappa^{2}}
-24H^{2}\alpha'\dot{\phi}\bigg)}\partial^{2}\Psi\,.
\end{equation}

By substituting the constraint (\ref{56}) into the action
(\ref{55}), we find
\begin{equation}
\label{58} S_{2}=\int dt\, d^{3}x\, a^{3}\,
{\cal{U}}\bigg[\dot{\Psi}-\frac{c_{s}^{2}}{a^{2}}\big(\partial\Psi\big)^{2}\bigg]\,,
\end{equation}
where
\begin{equation}
{\cal{U}}=\frac{\bigg(\frac{1}{\kappa^{2}}-8H\dot{\alpha}\bigg)\bigg(\Big(\frac{4}{\kappa^{2}}
-32H\dot{\alpha}\Big)\Big(144H^{3}\dot{\alpha}-\frac{9H^{2}}{\kappa^{2}}+\frac{3\dot{\phi}^{2}}{2\sqrt{1-f\dot{\phi}^{2}}}+
\frac{3f\,\dot{\phi}^{4}}{2\big(1-f\dot{\phi}^{2}\big)^{\frac{3}{2}}}\Big)+9\Big(\frac{2H}{\kappa^{2}}-24H^{2}\dot{\alpha}\Big)
\bigg)} {3\Big(\frac{2H}{\kappa^{2}}-24H\dot{\alpha}\Big)^{2}}\,,
\label{59}
\end{equation}
and
\begin{eqnarray}
c_{s}^{2}=\frac{3\bigg(\Big(\frac{2}{\kappa^{2}}-16H\dot{\alpha}\Big)\Big(\frac{2H}{\kappa^{2}}-24H^{2}\dot{\alpha}\Big)H
-\Big(\frac{8H}{\kappa^{2}}-96H^{2}\dot{\alpha}\Big)\Big(8\dot{H}\dot{\alpha}+8H\ddot{\alpha}\Big)\bigg)}
{\Big(\frac{4}{\kappa^{2}}-32H\dot{\alpha}\Big)\Big(144H^{3}\dot{\alpha}
-\frac{9H^{2}}{\kappa^{2}}+\frac{3\dot{\phi}^{2}}{2\sqrt{1-f\dot{\phi}^{2}}}+
\frac{3f\,\dot{\phi}^{4}}{2\big(1-f\dot{\phi}^{2}\big)^{\frac{3}{2}}}\Big)
+9\Big(\frac{2H}{\kappa^{2}}-24H\dot{\alpha}\Big)^{2}}\nonumber
\end{eqnarray}
\begin{equation}
-\frac{\frac{3\Big(\frac{2H}{\kappa^{2}}-24H^{2}\dot{\alpha}\Big)^{2}}{\frac{1}{\kappa^{2}}-8H\dot{\alpha}}
\Big(\frac{1}{\kappa^{2}}-8\ddot{\alpha}\Big)+\Big(\frac{2}{\kappa^{2}}-16H\dot{\alpha}\Big)
\Big(\frac{2\dot{H}}{\kappa^{2}}-48H\dot{H}\dot{\alpha}-24H^{2}\ddot{\alpha}\Big)}
{\Big(\frac{4}{\kappa^{2}}-32H\dot{\alpha}\Big)\Big(144H^{3}\dot{\alpha}
-\frac{9H^{2}}{\kappa^{2}}+\frac{3\dot{\phi}^{2}}{2\sqrt{1-f\dot{\phi}^{2}}}+
\frac{3f\,\dot{\phi}^{4}}{2\big(1-f\dot{\phi}^{2}\big)^{\frac{3}{2}}}\Big)
+9\Big(\frac{2H}{\kappa^{2}}-24H\dot{\alpha}\Big)^{2}} \,.\label{60}
\end{equation}
For more details to obtain the equations of this section, one can
refer to \cite{Fel11a,Fel11b,Che08,See05}. Since our aim in this
section is the study of the three-point correlation function of the
perturbations, we should expand the action (\ref{58}) up to the
third order. The explicit form of the third-order action is given in
the Appendix \textbf{A}. To proceed, we introduce the parameter
$\hat{B}$ which, by using equation (\ref{57}), relates two
perturbation parameters as follows
\begin{equation}
B=-\frac{2\kappa^{-2}-16H\dot{\alpha}}{2\kappa^{-2}H-24H^{2}\alpha'\dot{\phi}}+\hat{B}\,,\quad
\partial^{2}\hat{B}=\frac{a^{2}{\cal{U}}\dot{\Psi}}{\kappa^{-2}-8H\dot{\alpha}}\,. \label{61}
\end{equation}
On the other hand, the variation of the Lagrangian of action
(\ref{55}) with respect to $\Psi$, gives us the equation of motion
of $\Psi$ as follows
\begin{equation}
\frac{d}{dt}\Big(a^{3}{\cal{U}}\dot{\Psi}\Big)-a\,{\cal{U}}c_{s}^{2}\partial^{2}\Psi=0\,.
\label{62}
\end{equation}

Now, by using equations (\ref{56})-(\ref{62}), we can rewrite the
cubic action up to the leading order as follows
\begin{equation}
S_{3}=\int dt\, d^{3}x\,
\Bigg\{3a^{3}\Bigg[{\cal{U}}\bigg(1-\frac{1}{c_{s}^{2}}\bigg)\Bigg]
\Psi\dot{\Psi}^{2}+a\Bigg[c_{s}^{2}\,{\cal{U}}\Big(\frac{1}{c_{s}^{2}}-1\Big)\Bigg]
\Psi\Big(\partial\Psi\Big)^{2} +a^{3}\Bigg[\frac{{\cal{U}}}{\kappa
H}\Big(\frac{1}{c_{s}^{2}}-1-\frac{2\lambda}{\chi}\Big)\Bigg]
\dot{\Psi}^{3}\Bigg\}\,. \label{63}
\end{equation}
where the parameters $\lambda$ and $\chi$ are defined respectively
as
\begin{equation}
\lambda=\frac{f\,\dot{\phi}^{4}}{4\big(1-f\,\dot{\phi}^{2}\big)^{\frac{3}{2}}}
+\frac{\,f^{2}\,\dot{\phi}^{6}}{3\big(1-f\,\dot{\phi}^{2}\big)^{\frac{5}{2}}}\,,
\label{64}
\end{equation}
\begin{equation}
\chi=\frac{\kappa^{4}}{4}\,{\cal{U}}\,\bigg(2\kappa^{-2}\,H-24\,H^{2}\,\dot{\alpha}\bigg)^{2}\,.
\label{65}
\end{equation}

To calculate the three point correlation function we use the
interaction picture where $H_{int}$ is equal to ${\cal{L}}_{3}$ (the
lagrangian of the cubic action). The vacuum expectation value of
$\Psi$ for the three-point operator in the conformal time interval
between $\tau_{i}$ and $\tau_{f}$ ($i$ and $f$ denote the beginning
and end of the inflation respectively) is given by the following
expression
\begin{equation}
\langle \Psi(\textbf{k}_{1})\, \Psi (\textbf{k}_{2})\, \Psi
(\textbf{k}_{3})\rangle= -i
\int_{\tau_{i}}^{\tau_{f}}d\tau\,a\,\langle
0|[\Psi(\textbf{k}_{1})\,\Psi (\textbf{k}_{2})\,
\Psi(\textbf{k}_{3})\,,\, H_{int}]|0\rangle\,. \label{66}
\end{equation}

By solving the integral of equation (\ref{66}), we find the
three-point correlation function as follows \cite{Che08,See05,Che02}
\begin{equation}
\langle \Psi(\textbf{k}_{1})\, \Psi (\textbf{k}_{2})\, \Psi
(\textbf{k}_{3})\rangle=\big(2\pi\big)^{3}\delta^{3}\big(\textbf{k}_{1}+\textbf{k}_{2}+\textbf{k}_{3}\big){\cal{B}}_{\Psi}\,.
\label{67}
\end{equation}
where
\begin{equation}
{\cal{B}}_{\Psi}=\frac{H^{4}{\cal{G}}_{\Psi}\big(k_{1},k_{2},k_{3}\big)}
{4\,{\cal{U}}^{2}c_{s}^{6}\,\prod_{i=1}^{3}k_{i}^{3}}\,. \label{68}
\end{equation}
Note that, in solving the integral of equation (\ref{66}), the
coefficients in the bracket are considered as constant
\cite{Fel11a,Che08}. This is because, these coefficients would
varies slower than the scale factor. In equation (\ref{68}), the
parameter ${\cal{G}}_{\Psi}$ is given by the following expression
\begin{equation}
{\cal{G}}_{\Psi}=
\frac{3}{4}\bigg(1-\frac{1}{c_{s}^{2}}\bigg)\,{\cal{S}}_{1}
+\frac{1}{4}\bigg(1-\frac{1}{c_{s}^{2}}\bigg)\,{\cal{S}}_{2}
+\frac{3\kappa}{2}\Big(\frac{1}{c_{s}^{2}}-1-\frac{2\lambda}{\chi}\Big)
\,{\cal{S}}_{3}\,. \label{70}
\end{equation}
where $K$ is defined as $K=\sum_{i} k_{i}$ and the shape functions
${\cal{S}}_{1}$, ${\cal{S}}_{2}$ and ${\cal{S}}_{3}$ are defined
respectively as
\begin{equation}
{\cal{S}}_{1}=\frac{2}{K}\sum_{i>j}k_{i}^{2}k_{j}^{2}-
\frac{1}{K^{2}}\sum_{i\neq j}k_{i}^{2}k_{j}^{3}\, \label{70-1}
\end{equation}
\begin{equation}
{\cal{S}}_{2}=\frac{1}{2}\sum_{i}k_{i}^{3}
+\frac{2}{K}\sum_{i>j}k_{i}^{2}k_{j}^{2}-\frac{1}{K^{2}} \sum_{i\neq
j}k_{i}^{2}k_{j}^{3} \label{70-2}
\end{equation}
and
\begin{equation}
{\cal{S}}_{3}= \frac{\big(k_{1}k_{2}k_{3}\big)^{2}}{K^{3}}\,.
\label{70-3}
\end{equation}

There are different shapes, depending on the values of momenta. The
momenta form a triangle and each shape has a pick in a configuration
of triangle. A local shape \cite{Gan94,Ver00,Wan00,Kom01} has a peak
in the squeezed limit (i.e., the limit where the modulus of the
momenta approaches $k_{3}\ll k_{1}\simeq k_{2}$). Another shape
which is corresponding to the equilateral configuration
\cite{Bab04b}, has a peak at $k_{1}=k_{2}=k_{3}$. There is another
shape which is orthogonal \cite{Sen10} to equilateral one. A linear
combination of the equilateral and orthogonal templates gives a
shape which is corresponding to folded triangle \cite{Che07} and has
a pick in $k_{1}=2k_{2}=2k_{3}$. An orthogonal non-Gaussianity has a
signal with a positive peak at the equilateral configuration and a
negative peak at the folded configuration. There is a parameter
$f_{_{NL}}$ which is called ``nonlinearity parameter''
\cite{Wan00,Gan94,Kom01,Bab04b} and measures the amplitude of the
non-Gaussianity. This dimensionless parameter is defined by the
following relation
\begin{equation}
f_{_{NL}}=\frac{10}{3}\frac{{\cal{G}}_{\Psi}}{\sum_{i=1}^{3}k_{i}^{3}}\,.
\label{71}
\end{equation}

Now, we study the amplitude of the non-Gaussianity in the
equilateral and orthogonal configurations. To this end, we follow
\cite{Fel13} and introduce a shape ${\cal{S}}_{*}^{equil}$ as
\begin{equation}
{\cal{S}}_{*}^{equil}=-\frac{12}{13}\Big(3{\cal{S}}_{1}-{\cal{S}}_{2}\Big)\,.
\label{72}
\end{equation}
There is another shape which is exactly orthogonal to (\ref{72}) and
is defined as \cite{Fel13}
\begin{equation}
{\cal{S}}_{*}^{ortho}=\frac{12}{14-13\beta}\Big(\beta\big(3{\cal{S}}_{1}-{\cal{S}}_{2}\big)+3{\cal{S}}_{1}-{\cal{S}}_{2}\Big)\,,
\label{73}
\end{equation}
where $\beta\simeq 1.1967996$. So, we can express the leading-order
bispectrum (\ref{70}), in terms of the equilateral basis
${\cal{S}}_{*}^{equil}$ and the orthogonal basis
${\cal{S}}_{*}^{ortho}$, as
\begin{equation}
{\cal{G}}_{\Psi}={\cal{C}}_{1}\,{\cal{S}}_{*}^{equil} +
{\cal{C}}_{2} \,{\cal{S}}_{*}^{ortho}\,, \label{74}
\end{equation}
where
\begin{equation}
{\cal{C}}_{1}=\frac{13}{12}\Bigg[\frac{1}{24}\bigg(1-\frac{1}{c_{s}^{2}}\bigg)\bigg(2+3\beta\bigg)
+\frac{\lambda}{12\chi}\bigg(2-3\beta\bigg) \Bigg]\,, \label{75}
\end{equation}
and
\begin{equation}
{\cal{C}}_{2}=\frac{14-13\beta}{12}\Bigg[\frac{1}{8}\bigg(1-\frac{1}{c_{s}^{2}}\bigg)
-\frac{\lambda}{4\chi}\Bigg]\,. \label{76}
\end{equation}
In equations (\ref{75}) and (\ref{76}), $\lambda$ and $\chi$ are
defined by (\ref{64}) and (\ref{65}) respectively. By using
equations (\ref{71}) and (\ref{74}), we obtain
\begin{equation}
f_{_{NL}}^{equil}=\frac{130}{36\sum_{i=1}^{3}k_{i}^{3}}\Bigg[\frac{1}{24}\bigg(1-\frac{1}{c_{s}^{2}}\bigg)\bigg(2+3\beta\bigg)
+\frac{\lambda}{12\chi}\bigg(2-3\beta\bigg)
\Bigg]{\cal{S}}_{*}^{equil}\,, \label{77}
\end{equation}
and
\begin{equation}
f_{_{NL}}^{ortho}=\frac{140-130\beta}{36\,\sum_{i=1}^{3}k_{i}^{3}}\Bigg[\frac{1}{8}\bigg(1-\frac{1}{c_{s}^{2}}\bigg)
-\frac{\lambda}{4\chi}\Bigg]{\cal{S}}_{*}^{ortho}\,. \label{78}
\end{equation}

As we have stated previously, the shape function in the equilateral
configuration has a pick at $k_{1}=k_{2}=k_{3}$. Also an orthogonal
non-Gaussianity has a signal with a positive peak at the equilateral
configuration. So, equations (\ref{77}) and ({\ref{78}}) in the case
with $k_{1}=k_{2}=k_{3}$ give
\begin{equation}
f_{_{NL}}^{equil}=\frac{325}{18}\Bigg[\frac{1}{24}\bigg(\frac{1}{c_{s}^{2}}-1\bigg)\bigg(2+3\beta\bigg)
+\frac{\lambda}{12\chi}\bigg(2-3\beta\bigg) \Bigg]\,, \label{79}
\end{equation}
and
\begin{equation}
f_{_{NL}}^{ortho}=\frac{10}{9}\Big(\frac{65}{4}\beta+\frac{7}{6}\Big)\Bigg[\frac{1}{8}\bigg(1-\frac{1}{c_{s}^{2}}\bigg)
-\frac{\lambda}{4\chi}\Bigg]\,. \label{80}
\end{equation}

After obtaining the main equations of this setup, in the next
section we test this model in confrontation with recent
observational data to obtain some constraints on the model's
parameters. Our focus is mainly on the coefficient of the
Gauss-Bonnet term.

\section{Observational Constraints}
In the cosmological equations of this model, there are three
functions of the DBI field which have important role in the dynamics
of the model. These are $f(\phi)$, $V(\phi)$ and $\alpha(\phi)$.
Usually $f(\phi)$ is given in terms of the warp factor of the
AdS-like throat. In the pure AdS$_{5}$, $f(\phi)$ takes a simple
form as $f(\phi)=\beta\,\phi^{-4}$ \cite{Ali04}. On the other hand,
in reference \cite{Tsu13}, the authors have introduced another
function for $f(\phi)$ as $f(\phi)=\beta\,e^{\kappa\phi}$. So, we
first divide this section into two subsections; one with
$f(\phi)=\beta\phi^{4}$ and the other with $f(\phi)=\beta
e^{\kappa\phi}$. Then, we proceed our study by choosing the form of
the potential and the Gauss-Bonnet coupling term. Note that we take
the Guass-Bonnet coupling as
$\alpha(\phi)=\alpha_{_{GB}}{\cal{V}}(\phi)$. In constraining our
model with observational data, we focus mainly on $\alpha_{_{GB}}$.

\subsection{$f(\phi)=\beta\,\phi^{-4}$}
For this type of $f(\phi)$, we consider three types of potentials:
quadratic potential ($V=\frac{\sigma}{2}\phi^{2}$), quartic
potential ($V=\frac{\sigma}{4}\phi^{4}$) and exponential potential
($V(\phi)=\sigma\exp(-\kappa\phi)$). In which follows, we obtain
some constraints on the model parameters by analysis of these
parameters in the background of the Planck+WMAP9+ BAO data.

\subsubsection{$V(\phi)=\frac{\sigma}{2}\phi^{2}$}
In the first step, we consider a quadratic potential and adopt three
functions for ${\cal{V}}(\phi)$ as ${\cal{V}}(\phi)\sim\phi^{2}$,\,
$\phi^{4}$ and $e^{-\kappa\phi}$. With these choices, we solve the
integral of equation (\ref{21}). For ${\cal{V}}(\phi)\sim\phi^{2}$,
solving the integral gives
\begin{eqnarray}
N=-\frac {1}{64}\,\kappa^{2}\sigma\beta\,\ln
\left(-3\beta+8\kappa^{2}\alpha_{_{GB}}\,{\phi}^{2}\sigma
\beta+64\kappa^{2}
\alpha_{_{GB}}\,{\phi}^{4}+16\kappa^{2}{\alpha_{_{GB}}}^{2}
{\phi}^{2}R_{_{GB}}\beta \right)
-\frac{1}{32}\kappa^{4}\sigma^{2}\beta^{2}\alpha_{_{GB}}
\acute{A}\nonumber
\end{eqnarray}
\begin{equation}
-\frac{1}{16}\kappa^{4}\sigma\beta^{2}\alpha_{_{GB}}^{2}
R_{_{GB}}\,\acute{A}+\frac{1}{32}\kappa^{2}\sigma\beta\,\ln \left(
\sigma\,\beta+8\,{\phi}^{2}+2\,\alpha_{_{GB}}\, R_{_{GB}}\,\beta
\right)\, \Bigg|_{hc}^{f}\,, \label{81}
\end{equation}
where
\begin{equation}
\acute{A}=\frac{\arctan \left( \frac{1}{8}\frac
{8\kappa^{2}\alpha_{_{GB}}\,\sigma \beta+128\kappa^{2}
\alpha_{_{GB}}\,\phi^{2}+16\kappa^{2}\alpha_{_{GB}}^{2}
R_{_{GB}}\,\beta}{\sqrt
{12\kappa^{2}\alpha_{_{GB}}\,\beta+\kappa^{4}\alpha_{_{GB}}^{2}\sigma^{2}\beta^{2}+4\,{\kappa}^{4}
\alpha0_{_{GB}}^{3}\sigma\,\beta^{2}R_{_{GB}}+4 \kappa^{4}
\alpha_{_{GB}}^{4}R_{_{GB}}^{2}\beta^{2}}}
 \right) R_{_{GB}}}{{\sqrt {12\kappa^{2}\alpha_{_{GB}}\,
\beta+\kappa^{4}\alpha_{_{GB}}^{2}{\sigma}^{2}\beta^{2}+4\,
\kappa^{4}\alpha_{_{GB}}^{3}\sigma\,\beta^{2}R_{_{GB}}+4\kappa^{4}
\alpha_{_{GB}}^{4} R_{_{GB}}^{2}\beta^{2}}}}\,.
 \label{82}
\end{equation}
If we set Eq. (\ref{15}) equal to $1$ (corresponding to the end of
inflation), we obtain $\phi_{f}$. Then, by substituting this result
into equation (\ref{81}) we find $\phi_{hc}$. By substituting
$\phi_{hc}$ into the equations (\ref{47}) and (\ref{52}), we plot
the tensor to scalar ratio versus the spectral index (the left panel
of figure 1). The figure has been plotted for $N=50$ (the thinner
line) and $N=60$ (the thicker line) (this convention is applied
through this paper) in the background of the Planck+WMAP9+BAO joint
data. In all of the figures, the dashed lines are corresponding to
${\cal{V}}(\phi)\sim \phi^{2}$. As we see form this figure, for some
values of $\alpha_{_{GB}}$, the model is compatible with
observational data. In this case, for $N=50$, the model is
compatible with the joint $95\%$ CL of the Planck+WMAP9+BAO data if
$3\times 10^{-4}\leq\alpha_{_{GB}}<4.2\times10^{-3}$ . Also, for
$N=60$ this model is compatible with observational data if
$2.9\times 10^{-4}<\alpha_{_{GB}}<3.65\times 10^{-3}$. The right
panel of figure 1 shows the amplitude of the non-Gaussianity in the
orthogonal configuration versus the amplitude of the non-Gaussianity
in the equilateral configuration in the background of $68\%$, $95\%$
and $99\%$ CL of the Planck+WMAP9+BAO data. To plot this figure, we
substitute $\phi_{hc}$ in equations (\ref{79}) and (\ref{80}). Here
also, the figure has been plotted for $N=50$ (the thinner line) and
$N=60$ (the thicker line). For ${\cal{V}}(\phi)\sim \phi^{2}$ and
for $N=50$, the model is compatible with the joint $95\%$ CL of the
Planck+WMAP9+BAO data if $3.1\times
10^{-4}<\alpha_{_{GB}}<3.74\times 10^{-3}$. Also, for $N=60$ it is
compatible with observation, if $3\times
10^{-4}\leq\alpha_{_{GB}}\leq 3.96\times 10^{-3}$ . With this type
of ${\cal{V}}$, the model is well inside the $99\%$ CL of the
Planck+WMAP9+BAO data if $2.8\times
10^{-4}<\alpha_{_{GB}}<4.14\times 10^{-3}$ for $N=50$ and
$2.71\times 10^{-4}\leq\alpha_{_{GB}}\leq 4.22\times 10^{-3}$ for
$N=60$.

\begin{figure}[htp]
\begin{center}\includegraphics{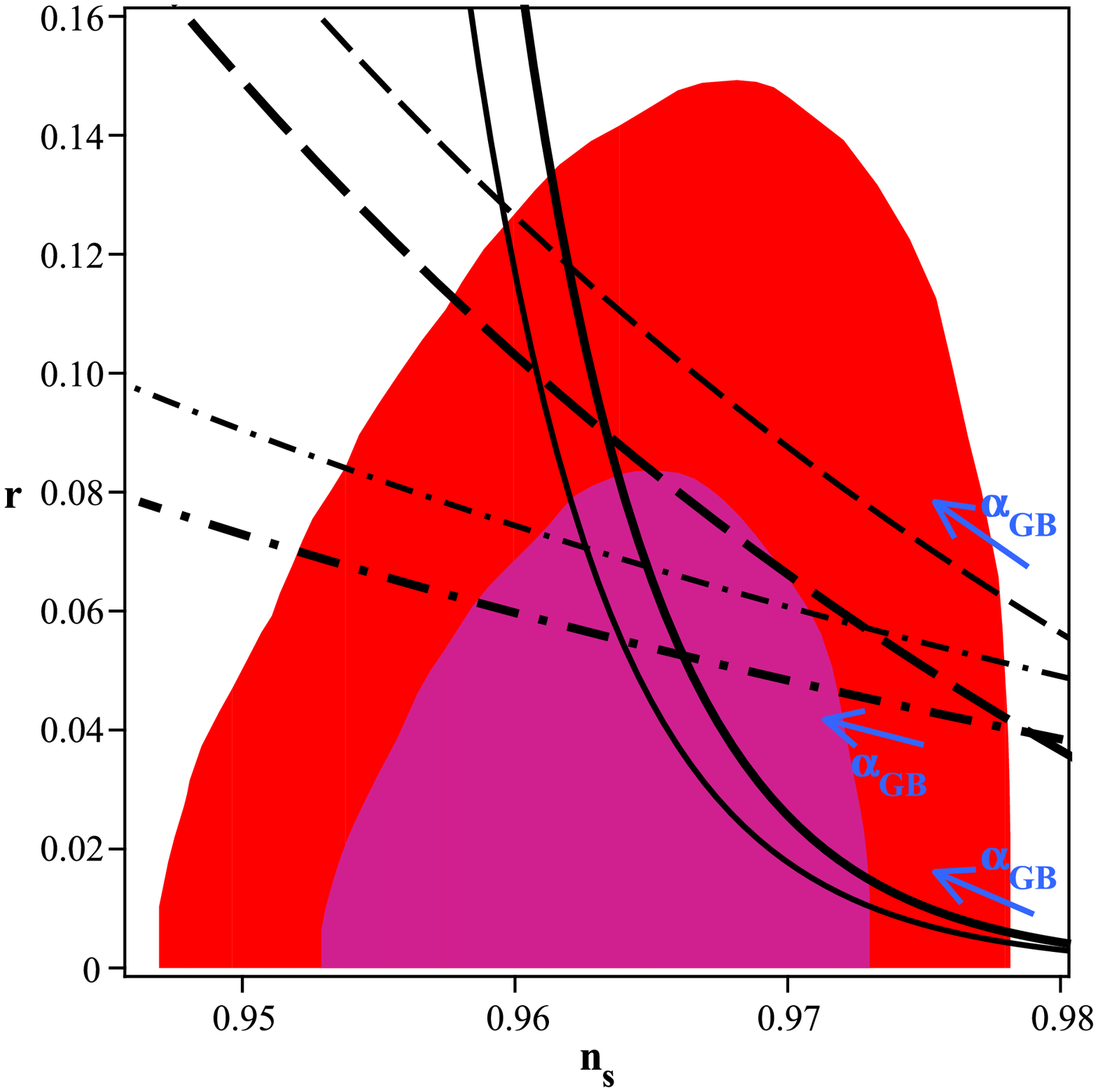} \hspace{1cm}
\includegraphics{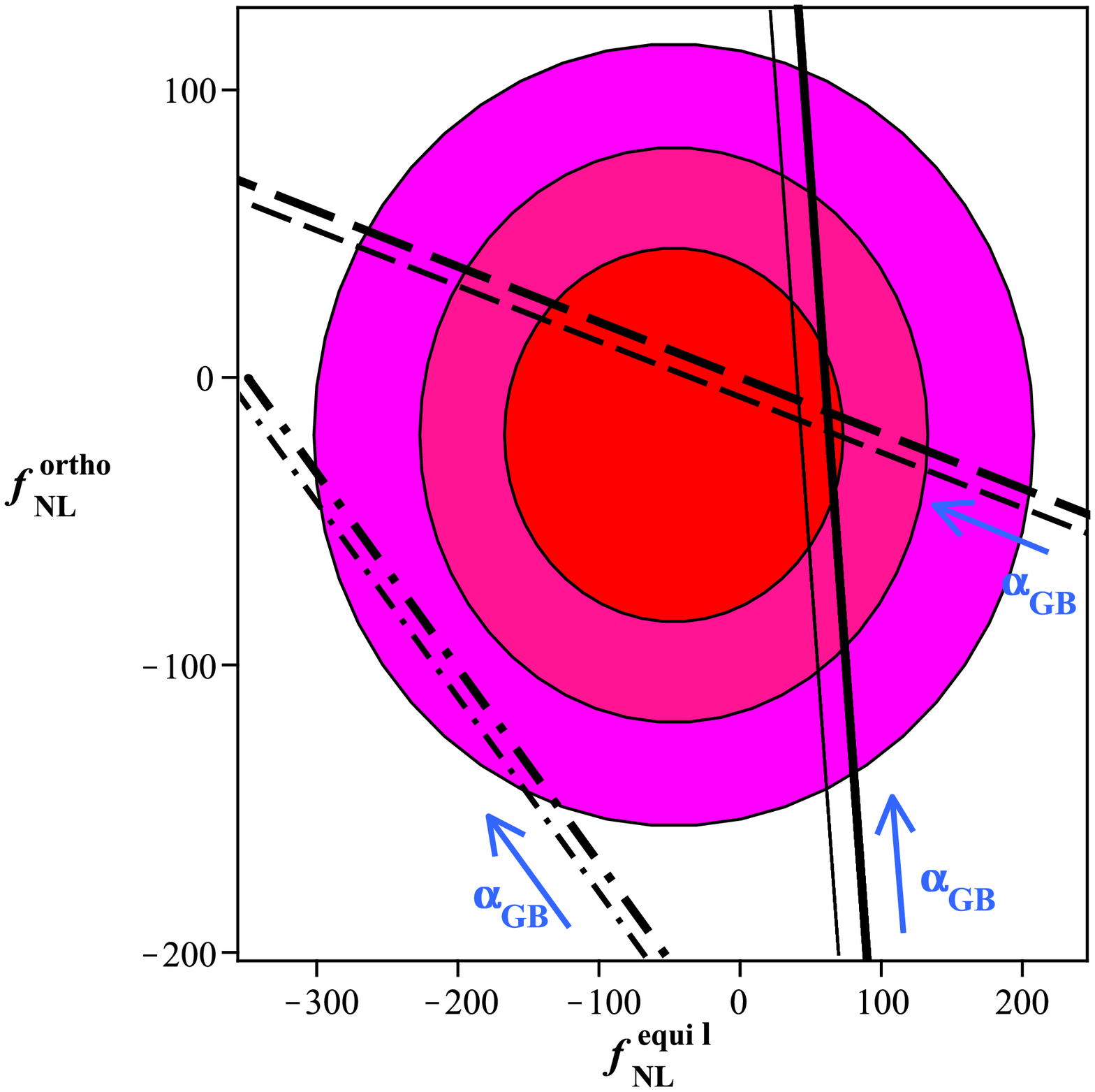} \vspace{6.5cm}
\end{center}
\caption{\small {Evolution of the tensor to scalar ratio versus the
spectral index (left panel) and the amplitude of the non-Gaussianity
in the orthogonal configuration versus the equilateral configuration
(right panel), for the case with $f(\phi)=\beta\phi^{-4}$ and with a
quadratic potential, in the background of the Planck+WMAP9+BAO data.
The figure has been plotted for $N=50$ (the thinner line) and $60$
(the thicker line). The solid lines are corresponding to
${\cal{V}}(\phi)\sim e^{-\kappa\phi}$, the dashed lines are
corresponding to ${\cal{V}}(\phi)\sim \phi^{2}$ and the dash-dotted
lines are corresponding to ${\cal{V}}(\phi) \sim \phi^{4}$ . For the
both values of $N$, a model with a non-minimal coupling between the
Gauss-Bonnet term and the DBI field, in some ranges of
$\alpha_{_{GB}}$ is compatible with observational data. }}
\end{figure}

Now, we adopt ${\cal{V}}(\phi)\sim \phi^{4}$ and solve the integral
of equation (\ref{21}). The result is given by the following
expression
\begin{equation}
\label{83} N=\frac{4}{9}\,\kappa^{4}\sigma\,
\alpha_{_{GB}}\,\phi^{6}+\frac{1}{4}\,{\frac {{\kappa
}^{2}\sigma\beta\,\ln
\left(\sigma\,\beta+8\phi^{2}+4\alpha_{_{GB}}\,\phi^{2}
R_{_{GB}}\,\beta\right) }{8+4\,\alpha_{_{GB}} \, R_{_{GB}}\,{\it
\beta}}}\, \Bigg|_{hc}^{f}\,,
\end{equation}
By finding $\phi_{hc}$ from this equation, we can plot the evolution
of the tensor to scalar ratio versus the spectral index (the
dash-dotted lines in the left panel of figure 1) by using of
equations (\ref{47}) and (\ref{52}). For ${\cal{V}}(\phi)\sim
\phi^{4}$ and for $N=50$, this model is inside the joint $95\%$ CL
of the Planck+WMAP9+BAO data if $6\times
10^{-4}<\alpha_{_{GB}}<2.44\times 10^{-3}$ . For $N=60$ this model
is compatible with observation if $6.23\times
10^{-4}<\alpha_{_{GB}}<2.524\times 10^{-3}$ . In studying the
amplitude of the non-Gaussianity we find that for this case, the
model both with $N=50$ and $N=60$ is outside the $95\%$ CL of the
Planck+WMAP9+BAO data. But, for $N=50$, the model with $5.2\times
10^{-4}<\alpha_{_{GB}}\leq 3\times 10^{-3}$ and for $N=60$, the
model with $5.5\times 10^{-4}<\alpha_{_{GB}}<3.02\times 10^{-3}$
lies inside the $99\%$ CL of the Planck+WMAP9+BAO data.

${\cal{V}}\sim e^{-\kappa\phi}$ is another case that we consider
here. By this function, solving the integral of equation (\ref{21})
gives
\begin{eqnarray}
N=\frac{1}{32}\,\sigma\,{\kappa}^{2}\beta\,\ln  \left(
\frac{1}{8}\,\sigma\,\beta+ \phi^{2} \right) +\frac{1}{2}\,{\frac
{\sigma\kappa^{2}
\left(\frac{8}{3}\,\alpha_{_{GB}}+\frac{8}{3}\,\kappa\,
\alpha_{_{GB}},\phi+\frac{4}{3}\kappa^{2}\alpha_{_{GB}}\,\phi^{2}
\right) }{e^{\kappa\phi}}}\nonumber
\end{eqnarray}
\begin{equation}
+\frac{1}{2}\,{\frac {\sigma\left( \kappa^{2}\phi^{2}{
e^{\kappa\,\phi}}-2\kappa\phi\,{ e^{\kappa\phi}}+2{e^{\kappa\phi}}
 \right) }{\kappa^{2}{\it R_{_{GB}}}\,\alpha_{_{GB}}}}+\frac{1}{4}\kappa^{2}{\phi
}^{2}-\acute{B}-{\frac {{\kappa}^{2}{\phi}^{4}}{\sigma\beta}}\,
\Bigg|_{hc}^{f}\,, \label{84}
\end{equation}
where
\begin{equation}
\acute{B}=32{\frac {\kappa^{6}{\phi}^{6}{ e^{\kappa\phi}}-6\kappa
^{5}{\phi}^{5}{ e^{\kappa\phi}}+30\kappa^{4}{\phi}^{4}{
e^{\kappa\phi}}-120\kappa^{3}{\phi}^{3}{ e^{\kappa
\phi}}+360\kappa^{2}{\phi}^{2}{e^{\kappa\phi}}-720\kappa\phi\,{
e^{\kappa\phi}}+720{ e^{\kappa\phi}}}{\kappa^{6}\sigma\beta^{2}
R_{_{GB}}\,\alpha_{_{GB}}}}\,. \label{85}
\end{equation}
The evolution of the tensor to scalar ratio versus the spectral
index is shown with the solid lines in the left panel of figure 1.
For ${\cal{V}}(\phi)\sim e^{-\kappa\phi}$ and for $N=50$, the model
is inside the joint $95\%$ CL of the Planck+WMAP9+BAO data if
$1.14\times 10^{-5}\leq\alpha_{_{GB}}\leq 4.5\times 10^{-3}$ . For
$N=60$ the constraint on the Gauss-Bonnet coupling parameter is as
$1.4\times 10^{-5}\leq\alpha_{_{GB}}\leq 4.4\times 10^{-3}$. In the
right panel of figure 1 we see the evolution of the amplitude of the
non-Gaussianity in the orthogonal configuration versus the
equilateral configuration in the background of $68\%$, $95\%$ and
$99\%$ CL of the Planck+WMAP9+BAO data. For ${\cal{V}}(\phi)\sim
e^{-\kappa\phi}$, the model is inside the joint $95\%$ CL of the
Planck+WMAP9+BAO data if $1.37\times 10^{-4}\leq\alpha_{_{GB}}\leq
4.61\times 10^{-3}$ for $N=50$ and $1.2\times
10^{-4}\leq\alpha_{_{GB}}\leq 4.581\times 10^{-3}$ for $N=60$. In
this case, the model lies inside the $99\%$ CL of the
Planck+WMAP9+BAO data if $1.03\times 10^{-4}<\alpha_{_{GB}}<
4.9\times 10^{-3}$ for $N=50$ and $1\times
10^{-4}\leq\alpha_{_{GB}}<5.21\times 10^{-3}$ for $N=60$.

\subsubsection{$V(\phi)=\frac{\sigma}{4}\phi^{4}$}
Now, we consider a qurtic potential and similar to the previous
subsection, we adopt three functions for ${\cal{V}}(\phi)$ as
${\cal{V}}(\phi)\sim\phi^{2}$,\, $\phi^{4}$ and $e^{-\kappa\phi}$.
At first, we solve the integral of equation (\ref{21}) for
${\cal{V}}(\phi)\sim\phi^{2}$ and the result is
\begin{eqnarray}
N=-\frac{3}{4}\,{\frac {{\kappa}^{2}\sigma\beta^{2} \alpha_{_{GB}}\,
R_{_{GB}} \,\ln \left(\sigma\phi^{2} \beta+8\phi^{2}+2\alpha_{_{GB}}
\, R_{_{GB}}\,\beta \right) }{ \left(3\sigma\beta+24 \right) \left(
\sigma\beta+8 \right) }}+24{\frac
{\kappa^{6}\sigma\beta^{3}\alpha_{_{GB}}^{4}R_{_{GB}}^{2}\acute{C}}{8
\kappa^{2}\alpha_{_{GB}}\,\sigma\beta+64\kappa^{2}\alpha_{_{GB}}
}}\nonumber
\end{eqnarray}
\begin{eqnarray}
+3{\frac{\kappa^{4}\sigma\beta^{2}\alpha_{_{GB}}^{2}R_{_{GB}}\,\ln
\left(-3\beta +8\kappa^{2}\beta\phi^{4}\sigma\beta+64\kappa^{2}
\alpha_{_{GB}}\phi^{4}+16\kappa^{2}\alpha_{_{GB}}^{2}{\phi}^{2}
R_{_{GB}}\,\beta\right) }{\left(3\sigma\beta+24 \right) \left(
8\kappa^{2}\alpha_{_{GB}}\,\sigma\beta+64\kappa^{2} \alpha_{_{GB}}
\right)}}+{\frac
{9}{16}}\kappa^{2}\sigma\beta^{2}\acute{C}\,\Bigg|_{hc}^{f}\,,
\label{86}
\end{eqnarray}
where
\begin{equation}
\acute{C}=\frac{\arctan\left( \frac{1}{4}{\frac{2\left(
8\kappa^{2}\alpha_{_{GB}}\,\sigma\beta+64\kappa^{2}\alpha_{_{GB}}\right)
\phi^ {2}+16\kappa^{2}\alpha_{_{GB}}^{2}R_{_{GB}}\,\beta}{\sqrt {6
\kappa^{2}\sigma\beta^{2}\alpha_{_{GB}}+48\alpha_{_{GB}}\,\beta\kappa^{2}+16\kappa^{4}
\alpha_{_{GB}}^{4}R_{_{GB}}^{2}\beta^{2}}}}\right)}{ \left(
3\sigma\beta+24 \right) { \sqrt {6\kappa^{2}\sigma\beta^{2}
\alpha_{_{GB}}+48\alpha_{_{GB}}\,\beta\,\kappa^{2}+16\kappa^{4}
\alpha_{_{GB}}^ {4}R_{_{GB}}^{2}\beta^{2}}}}\,, \label{87}
\end{equation}
By obtaining $\phi_{hc}$ and substituting it into equations
(\ref{47}) and (\ref{52}), we plot the evolution of the tensor to
scalar ratio versus the spectral index. The result is shown in the
left panel of figure 2. We have found that for ${\cal{V}}(\phi)\sim
\phi^{2}$, the model is compatible with the joint $95\%$ CL of the
Planck+WMAP9+BAO data if $1.5\times
10^{-4}\leq\alpha_{_{GB}}<6.2\times 10^{-3}$ and $1.71\times
10^{-4}<\alpha_{_{GB}}<6.34\times 10^{-3}$ for $N=50$ and $N=60$
respectively. The evolution of the amplitude of the non-Gaussianity
in the orthogonal configuration versus the equilateral configuration
in the background of $68\%$, $95\%$ and $99\%$ CL of the
Planck+WMAP9+BAO data is shown in the right panel of figure 2. For
${\cal{V}}(\phi)\sim \phi^{2}$ and for $N=50$, the model lies well
inside the joint $95\%$ CL of the Planck+WMAP9+BAO data if
$1.43\times 10^{-4}\leq\alpha_{_{GB}}<5.38\times 10^{-3}$ . Also,
for $N=60$ it is compatible with observation, if $1.6\times
10^{-4}<\alpha_{_{GB}}<6.014\times 10^{-3}$ . In this case, for
$N=50$ and $N=60$, the model lies within $99\%$ CL of the
Planck+WMAP9+BAO data if $1.3\times
10^{-4}\leq\alpha_{_{GB}}<5.8\times 10^{-3}$ and $1.52\times
10^{-4}<\alpha_{_{GB}}<6.4\times 10^{-3}$ respectively.

\begin{figure}[htp]
\begin{center}\includegraphics{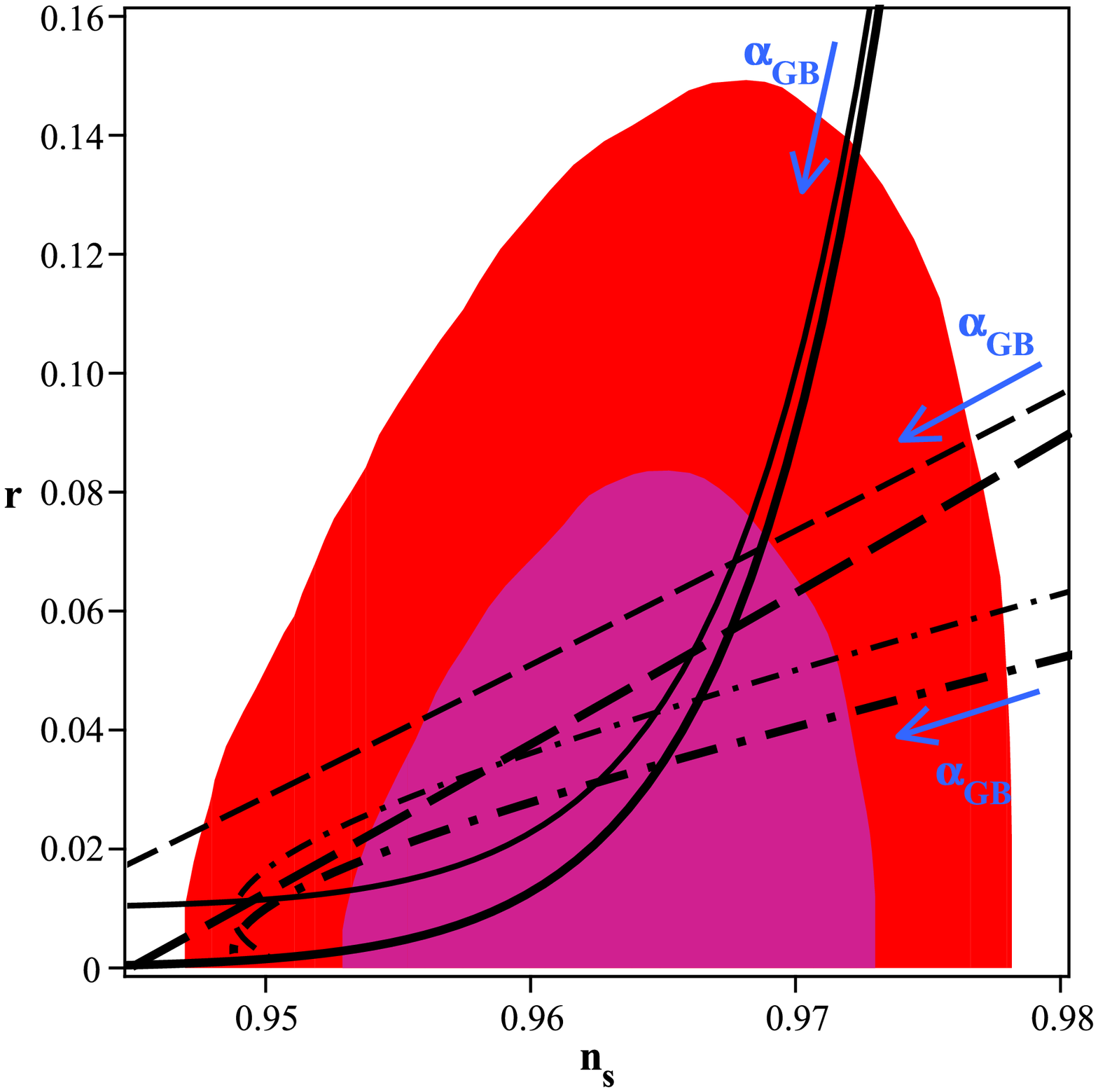} \hspace{1cm}
\includegraphics{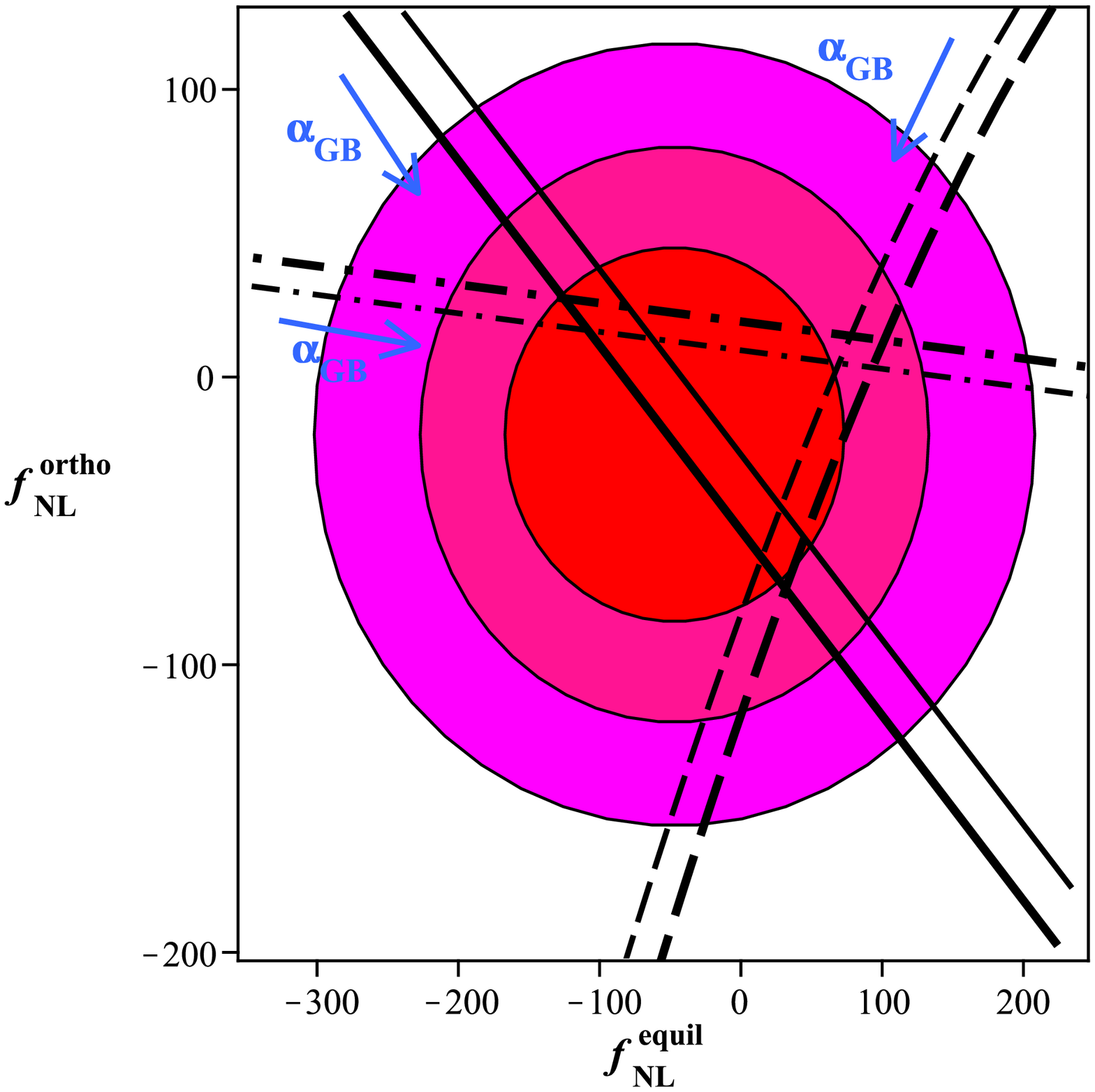} \vspace{6.5cm}
\end{center}
 \caption{\small {Evolution of the tensor to scalar ratio versus the
spectral index (left panel) and the amplitude of the non-Gaussianity
in the orthogonal configuration versus the equilateral configuration
(right panel), for the case with $f(\phi)=\beta\phi^{-4}$ and with a
quartic potential, in the background of Planck+WMAP9+BAO data. }}
\end{figure}

The result of solving the integral of equation (\ref{21}) with
${\cal{V}}(\phi)\sim \phi^{4}$ is given by
\begin{eqnarray}
N=\frac{\kappa^{2}\sigma\beta^{2}{3}^{\frac{5}{6}}{(32)}^{\frac{2}{3}}\arctan
\left( \frac{1}{3}\,\sqrt {3} \left(\frac{2}{3}\,{ \frac
{{3}^{\frac{2}{3}}\sqrt [3]{32}{\phi}^{2}}{\sqrt [3]{\acute{D}}}}+1
\right) \right)-{\kappa}^{2}\sigma\,{{\it f}}^{2}\sqrt
[3]{3}{(32)}^{\frac{2}{3}}\ln  \left( {\phi}^{2}-\frac{1}{32}\,\sqrt
[3]{3}{(32)}^{\frac{2}{3}}\sqrt [3]{\acute{D}} \right)}{12
{\acute{D}}^{\frac{5}{3}}}\nonumber
\end{eqnarray}
\begin{equation}
+\frac{1}{24}{\frac {\kappa^{2}\sigma\beta^{2} \sqrt
[3]{3}{(32)}^{\frac{2}{3}}\ln
\left(\phi^{4}+\frac{1}{32}\phi^{2}\sqrt [3]
{3}{(32)}^{\frac{2}{3}}\sqrt
[3]{\acute{D}}+\frac{1}{32}\,{3}^{\frac{2}{3}}\sqrt [
3]{32}{\acute{D}}^{\frac{2}{3}} \right)
}{\acute{D}\,{\acute{D}}^{2/3}}}\,\Bigg|_{hc}^{f}\,, \label{88}
\end{equation}
where
\begin{equation}
\acute{D}={\frac {\beta}{{\kappa}^{2}\alpha_{_{GB}}\, \left(
\sigma\beta+4 +2\alpha_{_{GB}}\,R_{_{GB}}\,\beta \right)}}
\label{89}
\end{equation}
For this case, the evolution of the tensor to scalar ratio versus
the scalar spectral index is shown with dash-dotted lines in the
left panel of figure 2. For ${\cal{V}}(\phi)\sim \phi^{4}$ and for
$N=50$, the model lies inside the joint $95\%$ CL of the
Planck+WMAP9+BAO data if $1.3 \times
10^{-5}<\alpha_{_{GB}}<8.6\times 10^{-4}$. Also, for $N=60$ the
model is compatible with observation if $1.38 \times
10^{-5}<\alpha_{_{GB}}<8.44\times 10^{-4}$. The evolution of the
$f^{ortho}$ versus $f^{equil}$ is shown in the right panel of figure
2. In studying the non-Gaussianity, it is obtained that, with this
function, the model with $1.65 \times 10^{-5}<\alpha_{_{GB}}\leq
8.471\times 10^{-4}$, and with $1.8 \times
10^{-5}<\alpha_{_{GB}}<8.479\times 10^{-4}$ for $N=60$ lies inside
the $95\%$ CL of the Planck+WMAP9+BAO data. Also, comparison with
the $99\%$ CL of the Planck+WMAP9+BAO data shows that $1.47 \times
10^{-5}<\alpha_{_{GB}}<8.77\times 10^{-4}$ for $N=50$, and $1.5
\times 10^{-5}<\alpha_{_{GB}}<8.8\times 10^{-4}$ for $N=60$.

By adopting ${\cal{V}}(\phi)\sim e^{-\kappa\phi}$, solving the
integral leads to
\begin{eqnarray}
N=\frac{1}{8}{\frac {\sigma\kappa^{2}{\phi}^{2}\beta}{\sigma\beta+
8}}+\frac{1}{3}{\frac {\sigma\alpha_{_{GB}} \left(
24+24\kappa\phi+12
\kappa^{2}{\phi}^{2}+4\kappa^{3}{\phi}^{3}+\kappa^{4}{\phi}^ {4}
\right) }{{ e^{\kappa\phi}}}}\nonumber
\end{eqnarray}
\begin{eqnarray}
+\frac{1}{4}\sigma\beta \Bigg( {\frac
{\sigma\left(\kappa^{4}\phi^{4}{
e^{\kappa\phi}}-4\kappa^{3}\phi^{3}{
e^{\kappa\phi}}+12\kappa^{2}\phi^{2}{
e^{\kappa\phi}}-24\kappa\phi{e^{\kappa\phi}}+24\,{{\rm
e}^{\kappa\,\phi}} \right) }{{\kappa}^{4}\alpha_{_{GB}}
R_{_{GB}}}}\nonumber
\end{eqnarray}
\begin{eqnarray}
+8{\frac {\kappa^{4}\phi^{4}{e^{
\kappa\phi}}-4\kappa^{3}\phi^{3}{e^{\kappa\phi}}+12
\kappa^{2}\phi^{2}{e^{\kappa\phi}}-24\kappa\phi{
e^{\kappa\phi}}+24{e^{\kappa\phi}}}{\kappa^{4}\alpha_{_{GB}}\,
R_{_{GB}}\,\beta}}+\frac{1}{2}\kappa^{2}{\phi}^{2} \Bigg) \left(
\sigma\beta+8 \right) ^{-1}\,\Bigg|_{hc}^{f}\,, \label{90}
\end{eqnarray}
The solid lines in the left panel of figure 2 show the evolution of
$r$ versus $n_{s}$. We have found that, for ${\cal{V}}(\phi)\sim
e^{-\kappa\phi}$ and for $N=50$, the model is well inside the joint
$95\%$ CL of the Planck+WMAP9+BAO data if $1.96\times
10^{-3}<\alpha_{_{GB}}<8.21\times 10^{-3}$ . For $N=60$, we find the
constraint on the Gauss-Bonnet coupling parameter as $1.85\times
10^{-3}<\alpha_{_{GB}}\leq 8.141\times 10^{-3}$. Also the solid
lines in the right panel of figure 2 show the evolution of
$f^{ortho}$ versus $f^{equil}$ for ${\cal{V}}(\phi)\sim
e^{-\kappa\phi}$ in the background of $68\%$, $95\%$ and $99\%$ CL
of the Planck+WMAP9+BAO data. With this function, the model is
compatible with the joint $95\%$ CL of the Planck+WMAP9+BAO data if
$2.05\times 10^{-3}\leq \alpha_{_{GB}}<7.1\times 10^{-3}$ for $N=50$
and $1.95\times 10^{-3}<\alpha_{_{GB}}\leq 7.66\times 10^{-3}$ for
$N=60$. Also, it is compatible with the joint $99\%$ CL of the
Planck+WMAP9+BAO data if $1.46\times
10^{-3}<\alpha_{_{GB}}<8.73\times 10^{-3}$ for $N=50$ and
$1.715\times 10^{-3}\leq\alpha_{_{GB}}\leq 7.9\times 10^{-3}$ for
$N=60$.

\subsubsection{$V(\phi)=\sigma e^{-\kappa\phi}$}
The third potential that we consider here, is an exponentially type
potential. With this potential, we solve the integral of equation
(\ref{21}) for ${\cal{V}}(\phi)\sim\phi^{2}$ to obtain
\begin{equation}
N=\frac{\sigma\kappa^{2}\left(-\frac{8}{3}\alpha_{_{GB}}-\frac{8}{3}
\alpha_{_{GB}}\kappa\phi \right)}{{ e^{\kappa\phi}}}+\phi+8\frac
{\kappa^{3}\phi^{3}-3\kappa^{2}\phi^{2}+6\kappa\phi-6}{{e^{-\kappa\phi}}\kappa^{4}\sigma\beta}+2{\frac{\alpha_{_{GB}}
R_{_{GB}}\left( \kappa\phi{e^{\kappa\phi}}-{e^{\kappa\phi}}\right)
}{\sigma\kappa^{2}}}\,\Bigg|_{hc}^{f}\,.\label{91}
\end{equation}
We find $\phi_{hc}$ from this equation and substitute it into the
equations (\ref{47}) and (\ref{52}) to obtain the evolution of
$n_{s}$ and $r$. One can see the evolution of the tensor to scalar
ratio versus the scalar spectral index in the left panel of figure 3
(the dashed lines). With ${\cal{V}}(\phi)\sim \phi^{2}$, the model
is compatible with the joint $95\%$ CL of the Planck+WMAP9+BAO data
if $8\times 10^{-5}<\alpha_{_{GB}}<3.3\times 10^{-4}$ and $4.8\times
10^{-4}<\alpha_{_{GB}}<1.73\times 10^{-3}$ for $N=50$ and
$8.36\times 10^{-5}<\alpha_{_{GB}}<3.84\times 10^{-4}$ and
$5.5\times 10^{-4}<\alpha_{_{GB}}<1.84\times 10^{-3}$ for $N=60$.
The evolution of the amplitude of the non-Gaussianity in the
orthogonal configuration versus the equilateral configuration in the
background of $68\%$, $95\%$ and $99\%$ CL of the Planck+WMAP9+BAO
data is shown in the right panel of the figure 3. In exploring the
amplitude of non-Gussianity we find that the model for $N=50$ lies
inside the joint $95\%$ CL of the Planck+WMAP9+BAO data if
$8.65\times 10^{-5}<\alpha_{_{GB}}\leq 1.643\times 10^{-3}$ . Also,
the model for $N=60$ lies inside the joint $95\%$ CL of the
Planck+WMAP9+BAO data if $7.9\times 10^{-5}<\alpha_{_{GB}}<1.5\times
10^{-3}$ . In this case, for $N=50$ and $N=60$, the model lies
within $99\%$ CL of the Planck+WMAP9+BAO data if $8\times
10^{-5}<\alpha_{_{GB}}<1.73\times 10^{-3}$ and $7.62\times
10^{-5}<\alpha_{_{GB}}<1.74\times 10^{-3}$ respectively.

\begin{figure}[htp]
\begin{center}\includegraphics{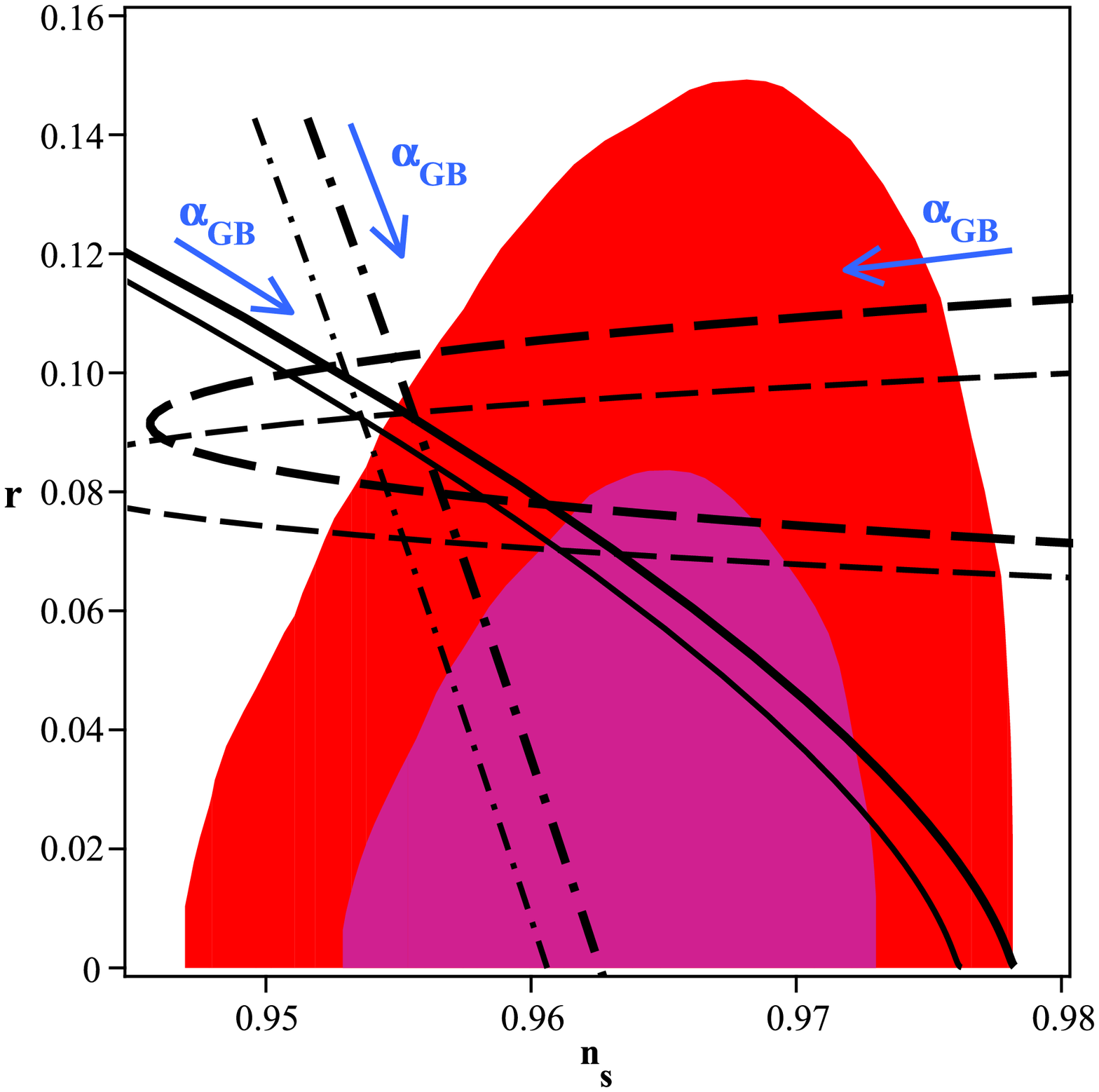} \hspace{1cm}
\includegraphics{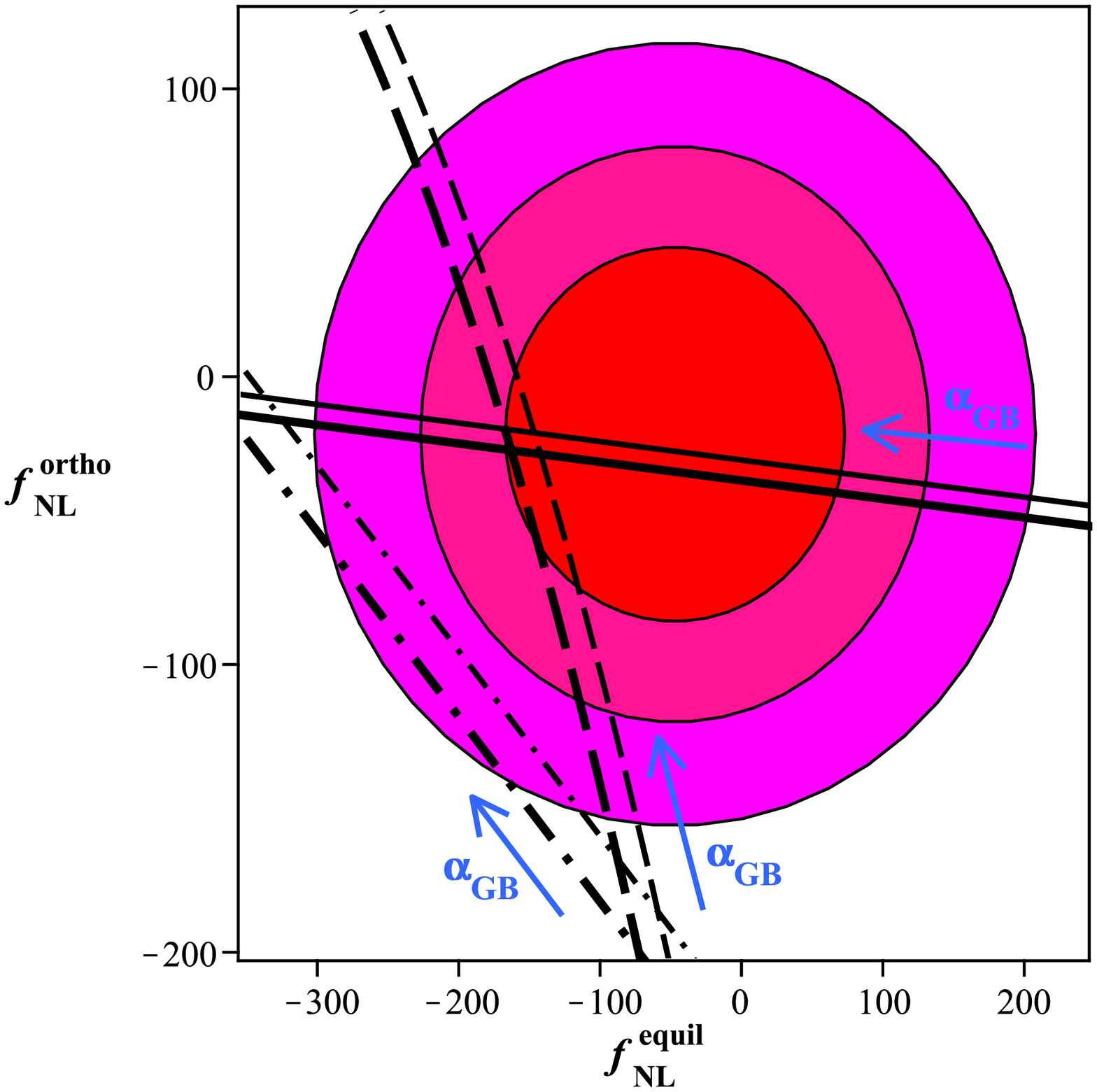} \vspace{6.5cm}
\end{center}
 \caption{\small { Evolution of the tensor to scalar ratio versus the
spectral index (left panel) and the amplitude of the non-Gaussianity
in the orthogonal configuration versus the equilateral
configuration(right panel), for the case with
$f(\phi)=\beta\phi^{-4}$ and with an exponential potential, in the
background of Planck+WMAP9+BAO data. }}
\end{figure}

The following equation is the result of  solving equation (\ref{21})
with ${\cal{V}}(\phi)\sim\phi^{4}$
\begin{eqnarray}
N=\frac{16}{3}{\frac {\sigma\alpha_{_{GB}}\left( 6+6\kappa\phi+3
\kappa^{2}\phi^{2}+\kappa^{3}\phi^{3}\right)}{{e^{
\kappa\phi}}}}+\phi+8{\frac {\kappa^{3}\phi^{3}{e^{
\kappa\phi}}-3\kappa^{2}\phi^{2}{e^{\kappa\phi}}+6 \kappa\phi{
e^{\kappa\phi}}-6{e^{\kappa\phi}}}{\kappa^{4}\sigma\beta}}\nonumber
\end{eqnarray}
\begin{equation}
+4{\frac {\alpha_{_{GB}}R_{_{GB}} \left( \kappa^{3}\phi^{3}{
e^{\kappa\phi}}-3\kappa^{2}\phi^{2}{
e^{\kappa\phi}}+6\kappa\phi{e^{\kappa\phi}}-6{ e^{\kappa\phi}}
\right) }{{\kappa}^{4}\sigma}}\,\Bigg|_{hc}^{f}\,,\label{92}
\end{equation}
One can see the evolution of $r$ versus $n_{s}$ in the left panel of
figure 3 (the dot-dashed lines). For ${\cal{V}}(\phi)\sim \phi^{4}$
and for $N=50$, the model lies inside the joint $95\%$ CL of the
Planck+WMAP9+BAO data if $7.5\times 10^{-5}<\alpha_{_{GB}}<6\times
10^{-3}$ . Also, for $N=60$ the model is compatible with
observation, if $7.66\times 10^{-5}<\alpha_{_{GB}}<6.33\times
10^{-3}$. The numerical study of the non-Gaussianity also, gives
some constraints on the $\alpha_{_{GB}}$ (see the dot-dashed lines
in the right panel of figure 3). With ${\cal{V}}(\phi)\sim
\phi^{4}$, the model for both $N=50$ and $N=60$ lies outside the
$95\%$ CL of the Planck+WMAP9+BAO data. But, it is compatible with
the $99\%$ CL of the Planck+WMAP9+BAO data if $7.38\times
10^{-5}<\alpha_{_{GB}}\leq5.16\times 10^{-3}$ for $N=50$, and
$5.66\times 10^{-5}<\alpha_{_{GB}}<5.4\times 10^{-3}$ for $N=60$.

Finally we consider an exponential function for ${\cal{V}}$ as
${\cal{V}}(\phi)\sim e^{-\kappa\phi}$ and solve the integral. We
obtain
\begin{equation}
N=\frac{2}{3}\,{\frac {\kappa^{4}\alpha_{_{GB}}\sigma}{{
e^{2\kappa\phi}}}}-\phi-8\,{\frac {\kappa^{3}\phi^{3}{
e^{\kappa\phi}}-3\kappa^{2}\phi^{2}{ e^{\kappa\phi}}+6\kappa\phi{
e^{\kappa\phi}}-6{ e^{\kappa\phi}}}{\sigma\kappa^ {4}\beta}}+{\frac
{\kappa\alpha_{_{GB}}R_{_{GB}}\,\phi}{\sigma}}\,\Bigg|_{hc}^{f}\,.\label{93}
\end{equation}
The solid lines in the left panel of figure 3 are corresponding to
${\cal{V}}(\phi)\sim e^{-\kappa\phi}$. With an exponential potential
and with ${\cal{V}}(\phi)\sim e^{-\kappa\phi}$, the model is well
inside the joint $95\%$ CL of the Planck+WMAP9+BAO data for $N=50$
and $N=60$, if $3.76\times 10^{-4}<\alpha_{_{GB}}<7.77\times
10^{-3}$ and $3.64\times 10^{-4}<\alpha_{_{GB}}<7.68\times 10^{-3}$
respectively. The solid lines in the right panel of figure 3 are
corresponding to ${\cal{V}}(\phi)\sim e^{-\kappa\phi}$. We have
found that with an exponential potential and with
${\cal{V}}(\phi)\sim e^{-\kappa\phi}$, the model is compatible with
the joint $95\%$ CL of the Planck+WMAP9+BAO data if $3.46\times
10^{-4}<\alpha_{_{GB}}<7.18\times 10^{-3}$ for $N=50$ and $3.4\times
10^{-4}<\alpha_{_{GB}}<7.26\times 10^{-3}$ for $N=60$. Also, a
comparison with the joint $99\%$ CL of the Planck+WMAP9+BAO data
shows that this model is compatible with observational data if
$3.261\times 10^{-4}\leq\alpha_{_{GB}}<7.29\times 10^{-3}$ for
$N=50$ and $3.13\times 10^{-4}<\alpha_{_{GB}}<7.5\times 10^{-3}$ for
$N=60$. The results of these arguments are summarized in tables 1
and 2.

\begin{table*}
\begin{tiny}
\caption{\label{tab:1} The ranges of $\alpha_{_{GB}}$ for which the
values of the inflationary parameters $r$ and $n_{s}$ are compatible
with the joint $95\%$ CL of the Planck+WMAP9+BAO data for
$f=\beta\phi^{-4}$}
\begin{tabular}{ccccc}
\\ \hline \hline$V$&$N$&${\cal{V}}(\phi)\sim \phi^{2}$&
${\cal{V}}(\phi)\sim \phi^{4}$&${\cal{V}}(\phi)\sim e^{-\kappa\phi}$\\ \hline\\
$\frac{\sigma}{2}\phi^{2}$& $N=50$& $3\times
10^{-4}\leq\alpha_{_{GB}}<4.2\times10^{-3}$ &$6\times
10^{-4}<\alpha_{_{GB}}<2.44\times 10^{-3}$&$1.14\times 10^{-5}\leq \alpha_{_{GB}}\leq4.5 \times 10^{-3}$\\\\
$\frac{\sigma}{2}\phi^{2}$& $N=60$&$2.9\times
10^{-4}<\alpha_{_{GB}}<3.65\times 10^{-3}$ &$6.23\times
10^{-4}<\alpha_{_{GB}}<2.524\times 10^{-3}$& $1.4\times 10^{-5}\leq \alpha_{_{GB}}\leq4.4\times 10^{-3}$\\\\
$\frac{\sigma}{4}\phi^{4}$& $N=50$ & $1.5\times
10^{-4}\leq\alpha_{_{GB}}<6.2\times10^{-3}$ & $1.3\times
10^{-5}\leq\alpha_{_{GB}}<8.6\times10^{-4}$ & $1.96\times
10^{-3}\leq\alpha_{_{GB}}<8.21\times10^{-3}$\\\\
$\frac{\sigma}{4}\phi^{4}$& $N=60$ & $1.71\times
10^{-4}\leq\alpha_{_{GB}}<6.34\times10^{-3}$ &$1.38\times
10^{-5}\leq\alpha_{_{GB}}<8.44\times10^{-4}$& $1.85\times
10^{-3}\leq\alpha_{_{GB}}<8.141\times10^{-3}$\\\\
$\sigma e^{-\kappa\phi}$& $N=50$ & $8\times
10^{-5}\leq\alpha_{_{GB}}<3.3\times10^{-4}$ & $7.5\times
10^{-5}\leq\alpha_{_{GB}}<6\times10^{-3}$ & $3.76\times
10^{-4}\leq\alpha_{_{GB}}<7.77\times10^{-3}$ \\
&&$4.8\times
10^{-4}\leq\alpha_{_{GB}}<1.73\times 10^{-3}$&&\\\\
$\sigma e^{-\kappa\phi}$& $N=60$ & $8.36\times
10^{-5}\leq\alpha_{_{GB}}<3.84\times10^{-4}$ & $7.66\times
10^{-5}\leq\alpha_{_{GB}}<6.33\times10^{-3}$ & $3.64\times
10^{-4}\leq\alpha_{_{GB}}<7.68\times10^{-3}$\\
&&$5.5\times
10^{-4}\leq\alpha_{_{GB}}<1.84\times10^{-3}$&&\\ \hline\\\\
\end{tabular}
\end{tiny}
\end{table*}
\begin{table*}
\begin{tiny}
\caption{\label{tab:2} The ranges of $\alpha_{_{GB}}$ for which the
values of the inflationary parameters $f^{ortho}$ and $f^{equil}$
are compatible with the joint $99\%$ CL of the Planck+WMAP9+BAO data
for $f=\beta\phi^{-4}$}
\begin{tabular}{ccccc}
\\ \hline \hline$V$& $N$ &${\cal{V}}(\phi)\sim \phi^{2}$&
${\cal{V}}(\phi)\sim \phi^{4}$&${\cal{V}}(\phi)\sim e^{-\kappa\phi}$\\ \hline\\
$\frac{\sigma}{2}\phi^{2}$& $N=50$ &$2.8\times
10^{-4}<\alpha_{_{GB}}<4.14\times 10^{-3}$  &$5.2\times
10^{-4}<\alpha_{_{GB}}\leq 3\times 10^{-3}$ & $1.03\times
10^{-4}<\alpha_{_{GB}}<
4.9\times 10^{-3}$ \\\\
$\frac{\sigma}{2}\phi^{2}$& $N=60$ & $2.71\times
10^{-4}\leq\alpha_{_{GB}}\leq 4.22\times 10^{-3}$ &$5.5\times
10^{-4}<\alpha_{_{GB}}<3.02\times 10^{-3}$ & $1\times
10^{-4}\leq\alpha_{_{GB}}<5.21\times 10^{-3}$ \\\\
$\frac{\sigma}{4}\phi^{4}$& $N=50$ & $1.3\times
10^{-4}\leq\alpha_{_{GB}}<5.8\times 10^{-3}$  & $1.47 \times
10^{-5}<\alpha_{_{GB}}<8.77\times 10^{-4}$ & $1.46\times
10^{-3}<\alpha_{_{GB}}<8.73\times 10^{-3}$ \\\\
$\frac{\sigma}{4}\phi^{4}$& $N=60$ &  $1.52\times
10^{-4}<\alpha_{_{GB}}<6.4\times 10^{-3}$ &$1.5
\times 10^{-5}<\alpha_{_{GB}}<8.8\times 10^{-4}$ & $1.715\times 10^{-3}\leq\alpha_{_{GB}}\leq 7.9\times 10^{-3}$\\\\
$\sigma e^{-\kappa\phi}$& $N=50$ & $8\times
10^{-5}<\alpha_{_{GB}}<1.73\times 10^{-3}$ & $7.38\times
10^{-5}<\alpha_{_{GB}}\leq5.16\times 10^{-3}$ & $3.261\times
10^{-4}\leq\alpha_{_{GB}}<7.29\times 10^{-3}$
\\\\
$\sigma e^{-\kappa\phi}$& $N=60$ & $7.62\times
10^{-5}<\alpha_{_{GB}}<1.74\times 10^{-3}$
& $5.66\times 10^{-5}<\alpha_{_{GB}}<5.4\times 10^{-3}$ & $3.13\times 10^{-4}<\alpha_{_{GB}}<7.5\times 10^{-3}$\\
&&  & &\\ \hline\\\\
\end{tabular}
\end{tiny}
\end{table*}

\subsection{$f(\phi)=\beta e^{\kappa\phi}$}
The second function we choose for $f(\phi)$ is an exponential
function. For this type of $f(\phi)$, similar to the previous part
we consider three types of potentials: quadratic, quartic and
exponential potential. Then, we obtain some constraints on the model
by comparing the main inflationary parameters of the model with the
Planck+WMAP9+ BAO data.

\subsubsection{$V(\phi)=\frac{\sigma}{2}\phi^{2}$}
With this potential, we consider three functions
${\cal{V}}(\phi)\sim\phi^{2}$,\, $\phi^{4}$ and $e^{-\kappa\phi}$
and we solve the integral of equation (\ref{21}). For
${\cal{V}}(\phi)\sim\phi^{2}$ we obtain
\begin{eqnarray}
N=\frac{1}{2}{\frac {\sigma\kappa^{2} \left(\frac{1}{4}\left(
8\sigma\kappa^{2}\alpha_{_{GB}}+16\kappa^{2}\alpha0_{_{GB}}^{2}
R_{_{GB}}\right) \phi^{4}+\frac{3}{2}\phi^{2} \right) }{6
\alpha_{_{GB}}R_{_{GB}}+3\sigma}}+ \frac{1}{4}{\frac
{\sigma^{2}\beta \left( \kappa^{2}\phi^{2}{
e^{\kappa\phi}}-2\kappa\phi{ e^{\kappa\phi}}+2{ e^{\kappa\phi}}
\right) }{\kappa^{2} \left( 2\alpha_{_{GB}}R_{_{GB}}+\sigma \right)
}}\nonumber
\end{eqnarray}
\begin{equation}
+\frac{1}{4}{ \frac{\sigma\kappa^{2}\phi^{2}}{2\alpha_{_{GB}}
R_{_{GB}}+\sigma}}+{ \frac {\sigma\alpha_{_{GB}}R_{_{GB}}\beta\left(
\kappa^{2}\phi^{2}{e^{\kappa\phi}}-2\kappa\phi{ e^{\kappa\phi}}+2 {
e^{\kappa\phi}} \right) }{2\kappa^{2} \left( 2\alpha_{_{GB}}
R_{_{GB}}+\sigma \right) }}\,\Bigg|_{hc}^{f}\,,\label{94}
\end{equation}
By finding $\phi_{hc}$ from this equation and then using equations
(\ref{47}) and (\ref{52}), we plot the evolution of the tensor to
scalar ratio versus the spectral index (left panel of figure 4). As
we see form this figure, for some values of $\alpha_{_{GB}}$, the
model is compatible with observational data. With
${\cal{V}}(\phi)\sim \phi^{2}$ and for $N=50$, the model is
observationally viable if $4.63\times
10^{-4}<\alpha_{_{GB}}<6.54\times 10^{-3}$ . For $N=60$ the model is
compatible with observational data if $4.47\times
10^{-4}<\alpha_{_{GB}}<6.4\times 10^{-3}$. The right panel of figure
4 shows the evolution of the amplitude of the non-Gaussianity in the
orthogonal configuration versus the equilateral configuration in the
background of $68\%$, $95\%$ and $99\%$ CL of the Planck+WMAP9+BAO
data. By adopting ${\cal{V}}(\phi)\sim \phi^{2}$, this model for
$N=50$ is compatible with the joint $95\%$ CL of the
Planck+WMAP9+BAO data if $4.713\times
10^{-4}\leq\alpha_{_{GB}}<6.7\times 10^{-3}$. Also, for $N=60$ it is
compatible with observation if $4.55\times
10^{-4}<\alpha_{_{GB}}<6.61\times 10^{-3}$. With this type of
${\cal{V}}$, the model is well inside the $99\%$ CL of the
Planck+WMAP9+BAO data if $4.66\times
10^{-4}<\alpha_{_{GB}}\leq6.774\times 10^{-3}$ for $N=50$ and
$4.41\times 10^{-4}<\alpha_{_{GB}}<6.73\times 10^{-3}$ for $N=60$.

\begin{figure}[htp]
\begin{center}\includegraphics{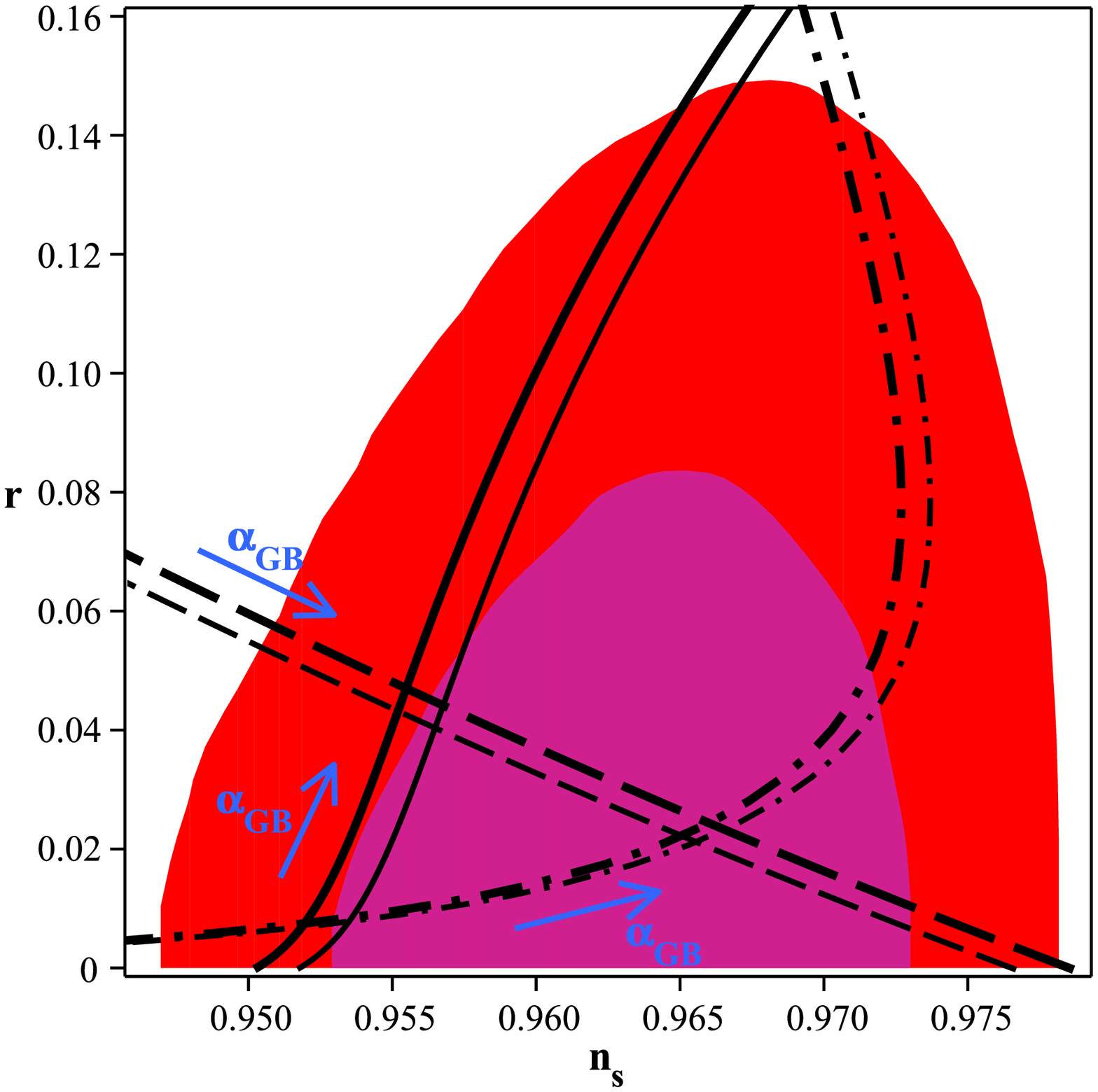} \hspace{1cm}
\includegraphics{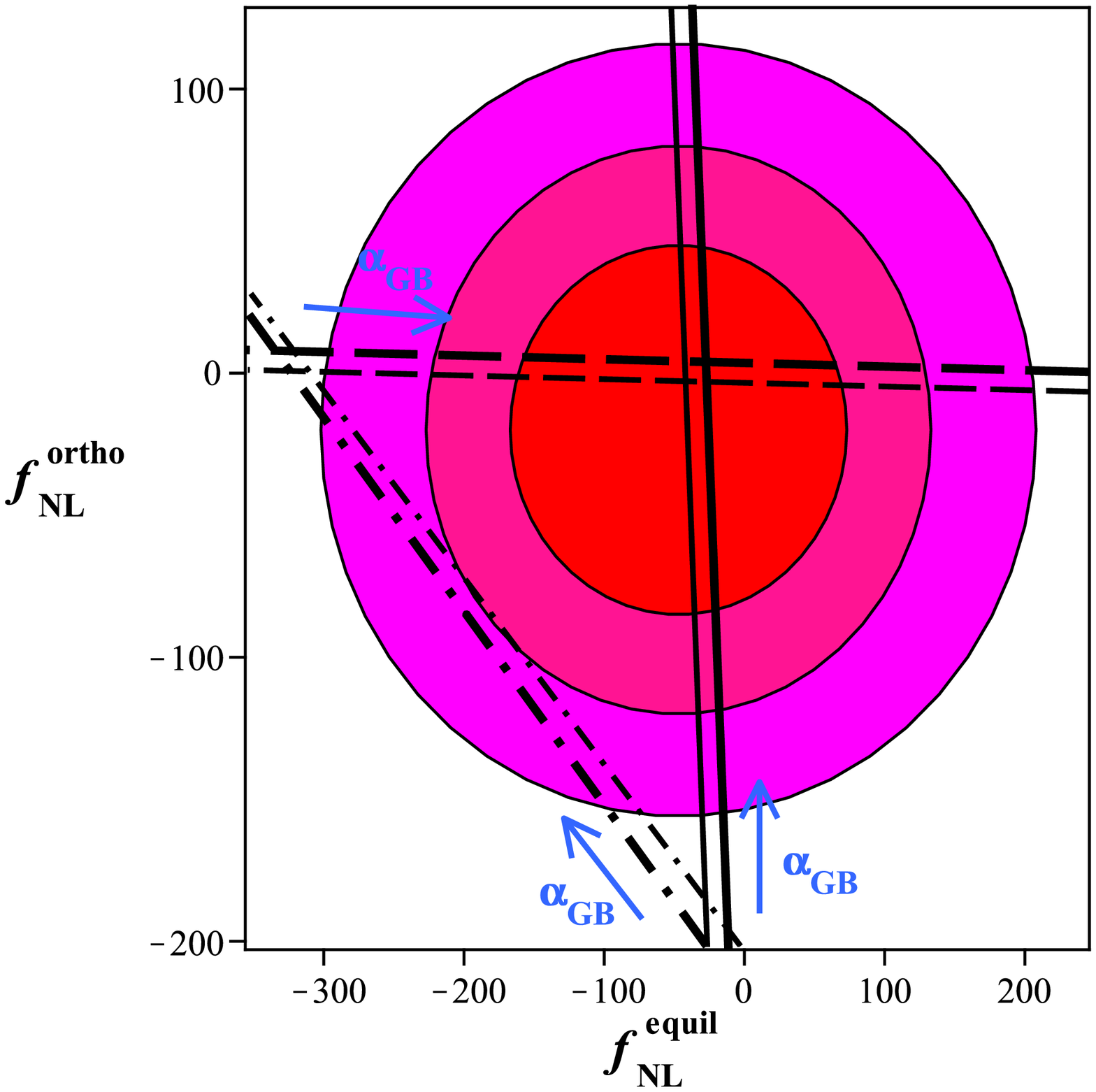} \vspace{6.5cm}
\end{center}
 \caption{\small {Evolution of the tensor to scalar ratio versus the
spectral index (left panel) and the amplitude of the non-Gaussianity
in the orthogonal configuration versus the equilateral configuration
(right panel), for the case with $f(\phi)=\beta e^{\kappa\phi}$ and
with a quadratic potential, in the background of the
Planck+WMAP9+BAO data. The figure has been plotted for $N=50$ (the
thinner line) and $70$ (the thicker line). The solid lines are
corresponding to ${\cal{V}}(\phi)\sim e^{-\kappa\phi}$, the dashed
lines are corresponding to ${\cal{V}}(\phi)\sim \phi^{2}$ and the
dash-dotted lines are corresponding to ${\cal{V}}(\phi) \sim
\phi^{4}$. For both values of $N$, a model with a non-minimal
coupling between the Gauss-Bonnet term and the DBI field, in some
ranges of $\alpha_{_{GB}}$ is compatible with observational data.
}}
\end{figure}

By solving the integral of equation (\ref{21}) with
${\cal{V}}(\phi)\sim \phi^{4}$, we obtain the following expression
\begin{eqnarray}
N=\frac{4}{9}\sigma\kappa^{4}\alpha_{_{GB}}\phi^{6}+\frac{1}{16}{\frac
{\sigma \kappa^{2}\ln \left( \sigma+4\alpha_{_{GB}}\phi^{2}R_{_{GB}}
\right)}{\alpha_{_{GB}}R_{_{GB}}}}+\frac{1}{16}{\frac{\kappa^{2}\ln
\left( \sigma\kappa^{2}+4\alpha_{_{GB}}\phi^{2}R_{_{GB}}\kappa^{2}
\right) }{\sigma\alpha_{_{GB}}R_{_{GB}}}}\nonumber
\end{eqnarray}
\begin{equation}
+\frac{1}{2}{\frac {\beta{
e^{\kappa\phi}}}{\sigma\kappa^{2}}}-\frac{1}{2}{\frac {\beta{
e^{\kappa\phi}}\phi}{\sigma\kappa}}+\frac{1}{4}{\frac {\beta{
e^{\kappa\phi}}{\phi}^{2}}{\sigma}}\,\Bigg|_{hc}^{f}\,.\label{95}
\end{equation}
With a quadratic potential, the model for $N=50$ is inside the joint
$95\%$ CL of the Planck+WMAP9+BAO data if $8.11\times
10^{-5}<\alpha_{_{GB}}< 3.42\times 10^{-2}$. Also for $N=60$ the
condition is $8\times 10^{-5}<\alpha_{_{GB}}<3.38\times 10^{-2}$
(see the left panel of figure 4). By treating the amplitude of the
non-Gaussianity (the right panel of figure 4) we find that, similar
to the case with $f(\phi)=\beta\,\phi^{-4}$, this model with a
quadratic potential and ${\cal{V}}(\phi)\sim \phi^{4}$, both for
$N=50$ and $N=60$ is outside the $95\%$ CL of the Planck+WMAP9+BAO
data. However, for $N=50$, the model with $3.1\times
10^{-4}<\alpha_{_{GB}}< 4.02\times 10^{-3}$ and for $N=60$, the
model with $3.18\times 10^{-4}<\alpha_{_{GB}}< 4.15\times 10^{-3}$
lies inside the $99\%$ CL of the Planck+WMAP9+BAO data.

The result of solving the integral of equation (\ref{21}) with
${\cal{V}}(\phi)\sim e^{-\kappa\phi}$ is as follows
\begin{eqnarray}
N=\frac{1}{4}\kappa^{2}\phi^{2}+\frac{1}{2}{\frac {\sigma\kappa^{2}
\left(\frac{8}{3}\alpha_{_{GB}}+\frac{8}{3}\kappa\phi
\alpha_{_{GB}}+\frac{4}{3}\kappa^{2}\alpha_{_{GB}}\phi^{2} \right)
}{{ e^{\kappa\phi}}}}\nonumber
\end{eqnarray}
\begin{equation}
-\frac{1}{4} \left( 2+\alpha_{_{GB}}R_{_{GB}}\beta
\right)\left(\frac{1}{2}{ \frac {\sigma\beta\left(
\kappa^{2}\phi^{2}{ e^{ \kappa\phi}}-2\kappa\phi{ e^{\kappa\phi}}+2{
e^{ \kappa\phi}} \right)
}{\kappa^{2}}}+\frac{1}{2}\kappa^{2}\phi^{2}-\frac{1}{
4}\kappa^{2}\alpha_{_{GB}}\phi^{2}R_{_{GB}}\beta\right)\,\Bigg|_{hc}^{f}\,.\label{96}
\end{equation}
The solid lines in the left pane of figure 4, that show the
evolution of $r$ versus $n_{s}$, are corresponding to
${\cal{V}}(\phi)\sim e^{-\kappa\phi}$. We have found that for
${\cal{V}}(\phi)\sim e^{-\kappa\phi}$ and for $N=50$, the model is
inside the joint $95\%$ CL of the Planck+WMAP9+BAO data if
$1.3\times 10^{-4}<\alpha_{_{GB}}<9.03\times 10^{-3}$ . For $N=60$,
we have found that the constraint on the Gauss-Bonnet coupling
parameter is as $1.41\times 10^{-4}<\alpha_{_{GB}}<9.1\times
10^{-3}$. Comparing the amplitude of the non-Gaussianity of this
model with observational data (the right panel of figure 4) shows
that, if we adopt ${\cal{V}}(\phi)\sim e^{-\kappa\phi}$, the model
is well inside the joint $95\%$ CL of the Planck+WMAP9+BAO data if
$1.64\times 10^{-4}<\alpha_{_{GB}}\leq 8.781\times 10^{-3}$ for
$N=50$ and $1.7\times 10^{-4}<\alpha_{_{GB}}<8.87\times 10^{-3}$ for
$N=60$. This model lies inside the $99\%$ CL of the Planck+WMAP9+BAO
data if $1.56\times 10^{-4}<\alpha_{_{GB}}<8.9\times 10^{-3}$ for
$N=50$ and $1.62\times 10^{-4}<\alpha_{_{GB}}<9\times 10^{-3}$ for
$N=60$.

\subsubsection{$V(\phi)=\frac{\sigma}{4}\phi^{4}$}
By considering a qurtic potential and adopting three functions for
${\cal{V}}(\phi)$, we solve the integral of equation (\ref{21}). The
result for ${\cal{V}}(\phi)\sim \phi^{2}$ is given by
\begin{eqnarray}
N=\frac{1}{36}{\frac{\kappa^{2} \left( 4\sigma^{2}\phi^{6}\kappa^{2
}\alpha_{_{GB}}+\frac{9}{2}\sigma\phi^{2} \right)
}{\sigma}}-\frac{1}{4}{\frac {\alpha_{_{GB}} R_{_{GB}}\kappa^{2}\ln
\left(\sigma\phi^{2}+2\alpha_{_{GB}}R_{_{GB}}\right)
}{\sigma}}+3{\frac
{\beta{e^{\kappa\phi}}\sigma}{\kappa^{4}}}-3{\frac {\beta{
e^{\kappa\phi}}\phi\sigma}{\kappa^{3}}}\nonumber
\end{eqnarray}
\begin{equation}
+\frac{3}{2}{\frac {\beta{
e^{\kappa\phi}}\phi^{2}\sigma}{{\kappa}^{2}}}-\frac{1}{2}{ \frac
{\beta{
e^{\kappa\phi}}\phi^{3}\sigma}{\kappa}}+\frac{1}{8}\sigma\beta{
e^{\kappa\phi}}\phi^{4}+\frac{1}{8}\kappa^{2}\phi^{2}-\frac{1}{4}{\frac
{\kappa\alpha_{_{GB}}R_{_{GB}}\ln \left(
\sigma\phi^{2}\kappa^{2}+2\alpha_{_{GB}} R_{_{GB}}\kappa^{2} \right)
}{\sigma}}\,\Bigg|_{hc}^{f}\,,\label{97}
\end{equation}
As usual, we obtain $\phi_{hc}$, substitute it into equations
(\ref{47}) and (\ref{52}) and then plot the evolution of the tensor
to scalar ratio versus the spectral index (left panel of figure 5).
With ${\cal{V}}(\phi)\sim \phi^{2}$, we find the constraint on
$\alpha_{_{GB}}$ as $3\times 10^{-3}<\alpha_{_{GB}}<4.5\times
10^{-2}$ (for $N=50$) and $3.08\times
10^{-3}<\alpha_{_{GB}}<4.44\times 10^{-2}$ (for $N=60$). By
comparing the non-Gaussianity of this model with observation (right
panel of figure 5) we see that in this case the model for $N=50$
lies well inside the joint $95\%$ CL of the Planck+WMAP9+BAO data if
$1.46\times 10^{-3}<\alpha_{_{GB}}\leq 4.143\times 10^{-2}$. Also,
for $N=60$ it is compatible with observation if $1.16\times
10^{-3}<\alpha_{_{GB}}\leq4.124\times 10^{-2}$. In this case, for
$N=50$ and $N=60$, the model lies within $99\%$ CL of the
Planck+WMAP9+BAO data if $1.3\times
10^{-3}<\alpha_{_{GB}}<4.26\times 10^{-2}$ and $1\times
10^{-3}<\alpha_{_{GB}}<4.38\times 10^{-2}$ respectively.

\begin{figure}[htp]
\begin{center}\includegraphics{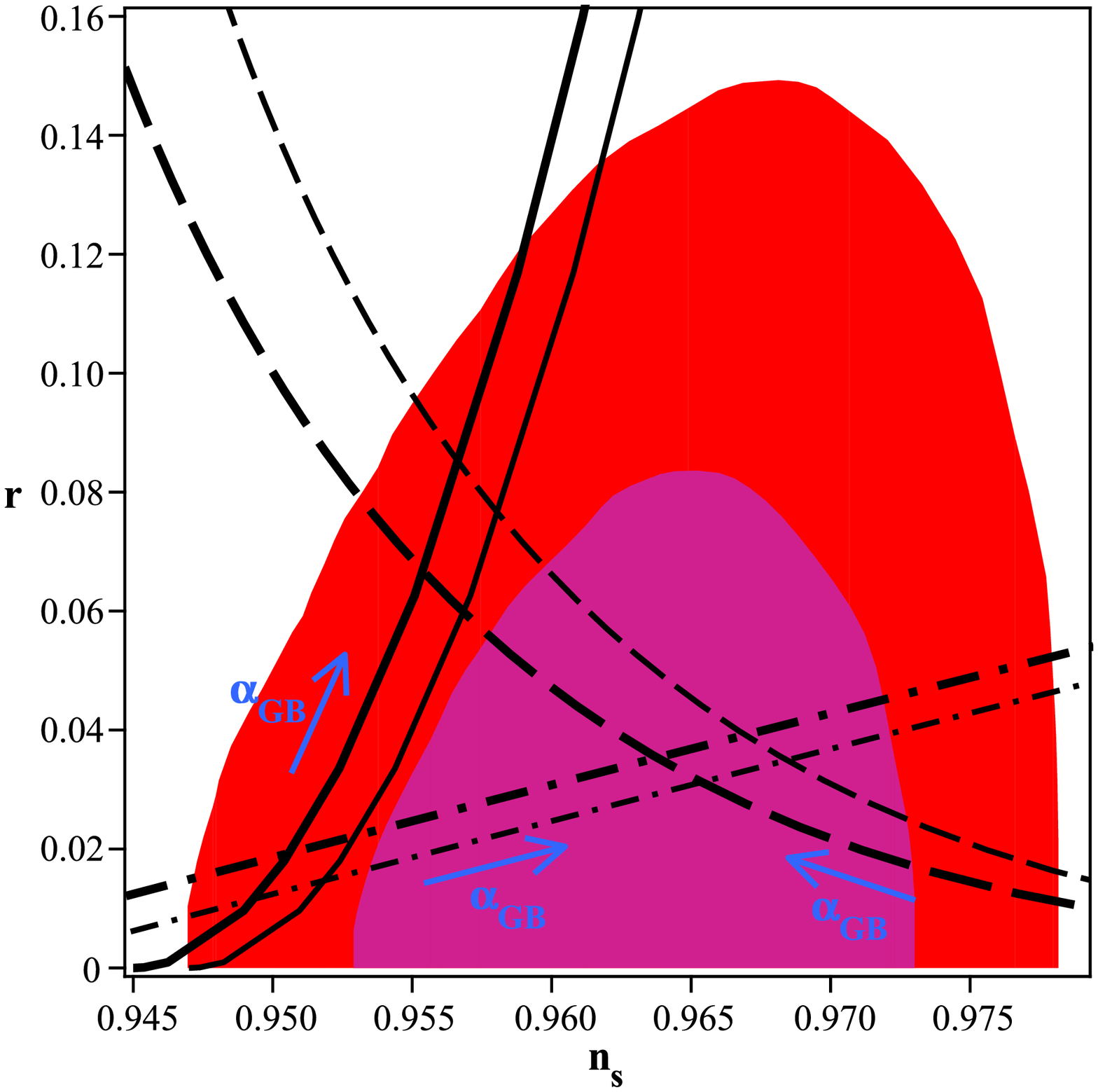} \hspace{1cm}
\includegraphics{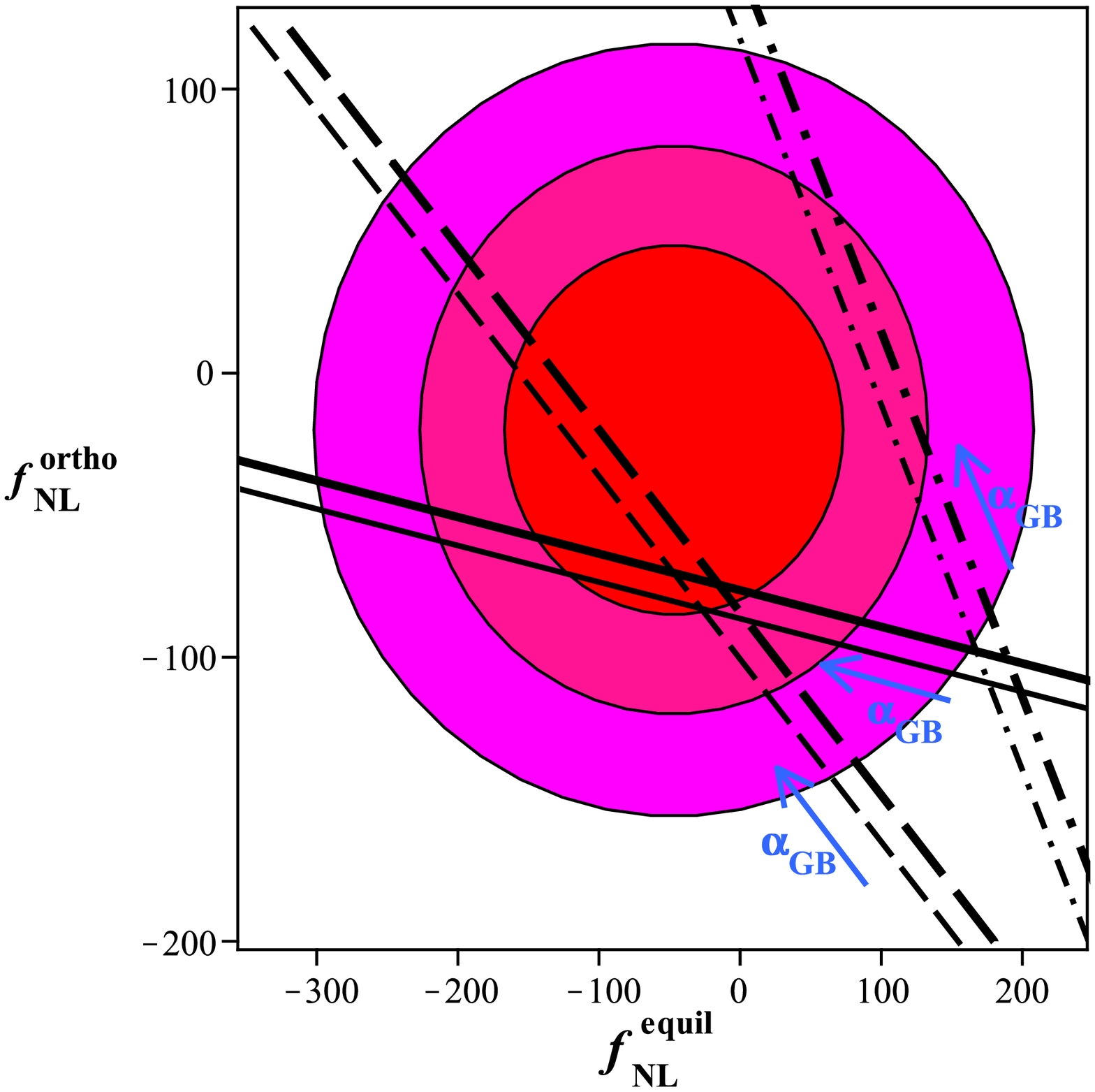} \vspace{6.5cm}
\end{center}
 \caption{\small {Evolution of the tensor to scalar ratio versus the
spectral index (left panel) and the amplitude of the non-Gaussianity
in the orthogonal configuration versus the equilateral configuration
(right panel), for the case with $f(\phi)=\beta e^{\kappa\phi}$ and
with a quartic potential, in the background of Planck+WMAP9+BAO
data.}}
\end{figure}

By adopting ${\cal{V}}(\phi)\sim \phi^{4}$, the integral (\ref{21})
gives
\begin{eqnarray}
N=\frac{1}{2}{\frac {\sigma\kappa^{2} \left( \frac{1}{8} \left(
32\sigma\kappa^{2}\alpha_{_{GB}}+64\alpha_{_{GB}}^{2}
R_{_{GB}}\kappa^{2}\right)\phi^{8}+\frac{3}{2}\phi^{2} \right)
}{6\sigma+12\alpha_{_{GB}} R_{_{GB}}}}+\frac{1}{4}{\frac
{\sigma\phi^{2}}{\left(\sigma+2\alpha_{_{GB}}R_{_{GB}} \right)
\kappa^{2}}}\nonumber
\end{eqnarray}
\begin{eqnarray}
+\frac{1}{16}{\frac {\sigma^{3}\beta\left( \kappa^{4}\phi^{4}{
e^{\kappa\phi}}-4\kappa^{3}\phi^{3 }{
e^{\kappa\phi}}+12\kappa^{2}\phi^{2}{e^{\kappa \phi}}-24\kappa\phi{
e^{\kappa\phi}}+24{ e^{\kappa\phi}} \right) }{ \left( \sigma+2
\alpha_{_{GB}}R_{_{GB}}\right) {\kappa}^{4}}}\nonumber
\end{eqnarray}
\begin{equation}
+\frac{1}{2}{\frac{\sigma\alpha_{_{GB}}R_{_{GB}}\beta\left(
\kappa^{4}\phi^{4}{ e^{\kappa\phi}}-4\kappa^{3}\phi^{3}{
e^{\kappa\phi}}+12\kappa^{2}\phi^{2}{ e^{\kappa\phi}}-24\kappa\phi{
e^{\kappa\phi}}+24{ e^{\kappa\phi}} \right) }{\left( \sigma+2
\alpha_{_{GB}} R_{_{GB}} \right)
{\kappa}^{4}}}\,\Bigg|_{hc}^{f}\,,\label{98}
\end{equation}
The dot-dashed lines in the left panel of figure 5 show the
evolution of $r$ versus $n_{s}$ with ${\cal{V}}(\phi)\sim \phi^{4}$.
This model for $N=50$ lies inside the joint $95\%$ CL of the
Planck+WMAP9+BAO data if $0.9\times 10^{-4}<\alpha_{_{GB}}<8.5\times
10^{-3}$ . Also, for $N=60$ the model is compatible with observation
if $0.83\times 10^{-4}<\alpha_{_{GB}}<8.38 \times 10^{-3}$. Other
constraints on $\alpha_{_{GB}}$ come from the study of the amplitude
of the non-Gaussianity. A comparison with the $95\%$ CL of the
Planck+WMAP9+BAO data shows that with this type of ${\cal{V}}$, the
model with $1.56\times 10^{-4}<\alpha_{_{GB}}<7.89\times 10^{-3}$,
for $N=50$, and with $1.3\times 10^{-4}<\alpha_{_{GB}}<8.1\times
10^{-3}$ for $N=60$ is compatible with observation. This model is
inside the $99\%$ CL of the Planck+WMAP9+BAO data if $1.43\times
10^{-4}<\alpha_{_{GB}}<8.02\times 10^{-3}$ for $N=50$, and
$1.15\times 10^{-4}<\alpha_{_{GB}}<8.26 \times 10^{-3}$ for $N=60$.

Similarly, the number of e-fold parameter with a quartic potential
and with ${\cal{V}}(\phi)\sim e^{-\kappa\phi}$ becomes
\begin{eqnarray}
N=\frac{1}{8}\phi^{2}\kappa^{2}+\frac{1}{3}{\frac{\sigma
\alpha_{_{GB}} \left(
24+24\kappa\phi+12\phi^{2}\kappa^{2}+4\kappa^{3}
\phi^{3}+\phi^{4}\kappa^{4} \right) }{{ e^{\kappa\phi}}}}
-\frac{1}{8} \left( 2+\alpha_{_{GB}}R_{_{GB}}\beta \right)\times
\nonumber
\end{eqnarray}
\begin{equation}
\left( -\frac{1}{2}{\frac {\sigma\beta \left( \phi^{4}\kappa^{4}{
e^{ \kappa\phi}}-4\kappa^{3}\phi^{3}{
e^{\kappa\phi}}+12\phi^{2}\kappa^{2}{ e^{\kappa\phi}}-24\kappa\phi{
e^{\kappa\phi}}+24{ e^{\kappa\phi}} \right) }{{\kappa
}^{4}}}+\frac{1}{2}\phi^{2}\kappa^{2}+\frac{1}{4}\phi^{2}\kappa^{2}
\alpha_{_{GB}} R_{_{GB}}\beta \right)\,\Bigg|_{hc}^{f}\,,\label{99}
\end{equation}
By studying the evolution of the tensor to scalar ratio versus the
spectral index (the solid lines in the left panel of figure 5) we
find that for ${\cal{V}}(\phi)\sim e^{-\kappa\phi}$ and for $N=50$,
the model is well inside the joint $95\%$ CL of the Planck+WMAP9+BAO
data if $6.86\times 10^{-4}<\alpha_{_{GB}}<6.44\times 10^{-3}$. For
$N=60$, we have found the constraint on the Gauss-Bonnet coupling
parameter as $6.67\times 10^{-4}<\alpha_{_{GB}}<6.31\times 10^{-3}$.
Also, studying the evolution of the amplitude of the non-Gaussianity
in the orthogonal configuration versus the amplitude of the
non-Gaussianity in the equilateral configuration (the solid lines in
the right panel of figure 5) shows that with quartic potential and
with ${\cal{V}}(\phi)\sim e^{-\kappa\phi}$, the model is compatible
with the joint $95\%$ CL of the Planck+WMAP9+BAO data if $6.71\times
10^{-4}<\alpha_{_{GB}}\leq 8.144\times 10^{-3}$ for $N=50$ and
$6.64\times 10^{-4}<\alpha_{_{GB}}<6.2\times 10^{-3}$ for $N=60$.
Also, it is compatible with the joint $99\%$ CL of the
Planck+WMAP9+BAO data if $6.62\times
10^{-4}<\alpha_{_{GB}}<8.3\times 10^{-3}$ for $N=50$ and $6.6\times
10^{-4}<\alpha_{_{GB}}\leq6.265\times 10^{-3}$ for $N=60$.

\subsubsection{$V(\phi)=\sigma e^{-\kappa\phi}$}
The last potential which we consider, is an exponential type
potential. By solving the integral of equation (\ref{21}) with this
type of potential and with ${\cal{V}}(\phi)\sim \phi^{2}$, we find
\begin{eqnarray}
N={\frac {\sigma\kappa^{2} \left( -\frac{8}{3}
\alpha_{_{GB}}-\frac{8}{3}\kappa\alpha_{_{GB}}\phi \right) }{{
e^{\kappa\phi}}}}+\sigma\kappa^{2}\beta\left(\phi-2{\frac
{\phi}{\sigma\beta}}+2{\frac {\alpha_{_{GB}} R_{_{GB}}\left(
\kappa\phi{ e^{\kappa\phi}}-{ e^{\kappa\phi}} \right)
}{\kappa^{3}\sigma}} \right)\,\Bigg|_{hc}^{f}\,,\label{100}
\end{eqnarray}
By plotting the evolution of the tensor to scalar ratio versus the
spectral index (the dashed lines in the left panel of figure 6) we
can find new constraints on $\alpha_{_{GB}}$. With an exponential
potential and with ${\cal{V}}(\phi)\sim \phi^{2}$, the model is
compatible with observation if $6.1\times
10^{-4}<\alpha_{_{GB}}<8.9\times 10^{-3}$ for $N=50$ and $5.97\times
10^{-4}<\alpha_{_{GB}}<8.81\times 10^{-3}$ for $N=60$. Also by
considering the amplitude of non-Gaussianity, we find other
constraints on $\alpha_{_{GB}}$ (see the dashed lines in the right
panel of figure 6). In this case, the model for $N=50$ lies inside
the joint $95\%$ CL of the Planck+WMAP9+BAO data if $6.44\times
10^{-4}<\alpha_{_{GB}}<9.27\times 10^{-3}$. Also, the model for
$N=60$ lies inside the joint $95\%$ CL of the Planck+WMAP9+BAO data
if $6.3\times 10^{-4}<\alpha_{_{GB}}<9.02\times 10^{-3}$. For $N=50$
and $N=60$, the model lies within $99\%$ CL of the Planck+WMAP9+BAO
data if $6.23\times 10^{-4}<\alpha_{_{GB}}<9.36\times 10^{-3}$ and
$6.125\times 10^{-4}\leq\alpha_{_{GB}}\leq9.141\times 10^{-3}$
respectively.

\begin{figure}[htp]
\begin{center}\includegraphics{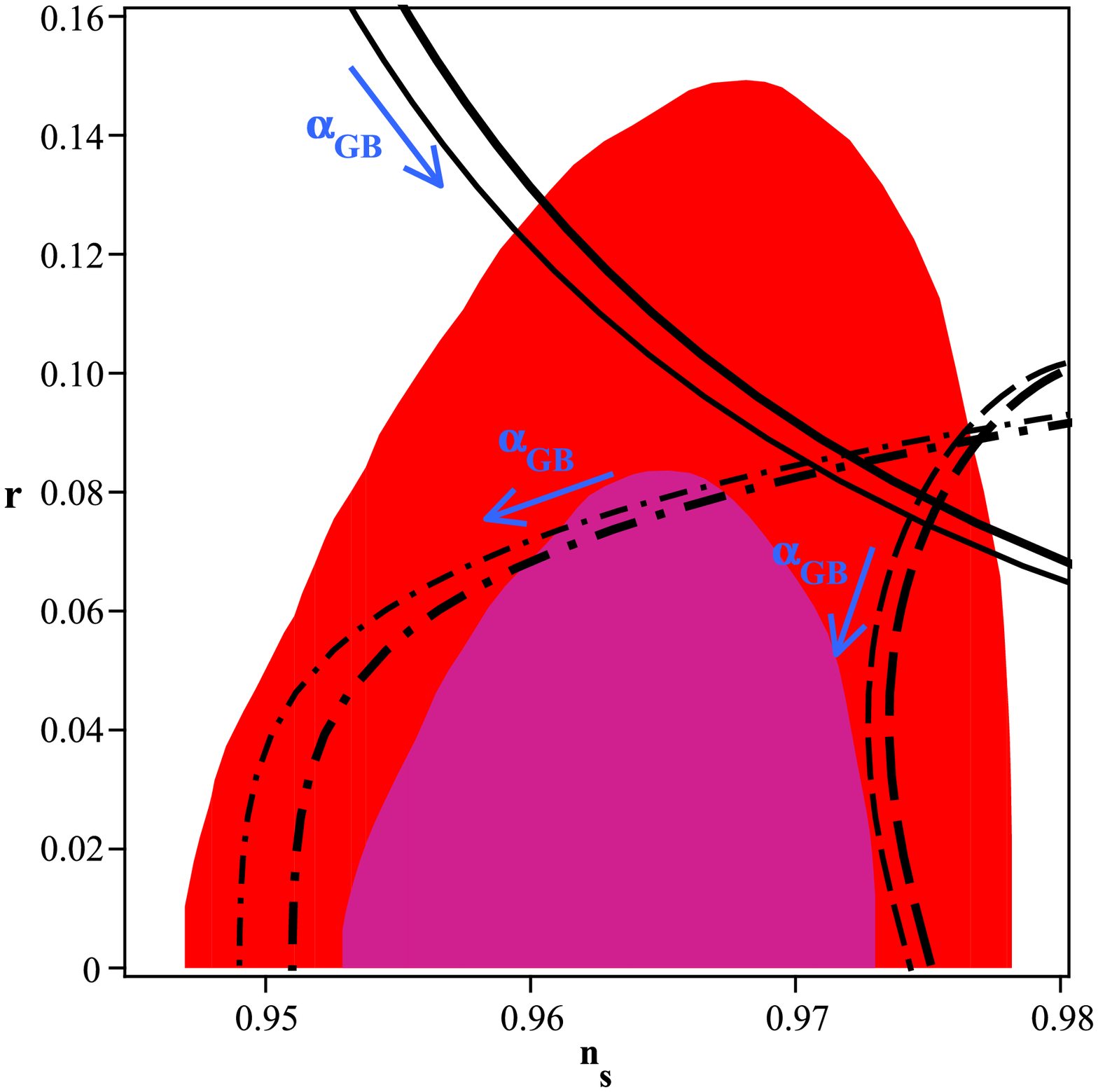} \hspace{1cm}
\includegraphics{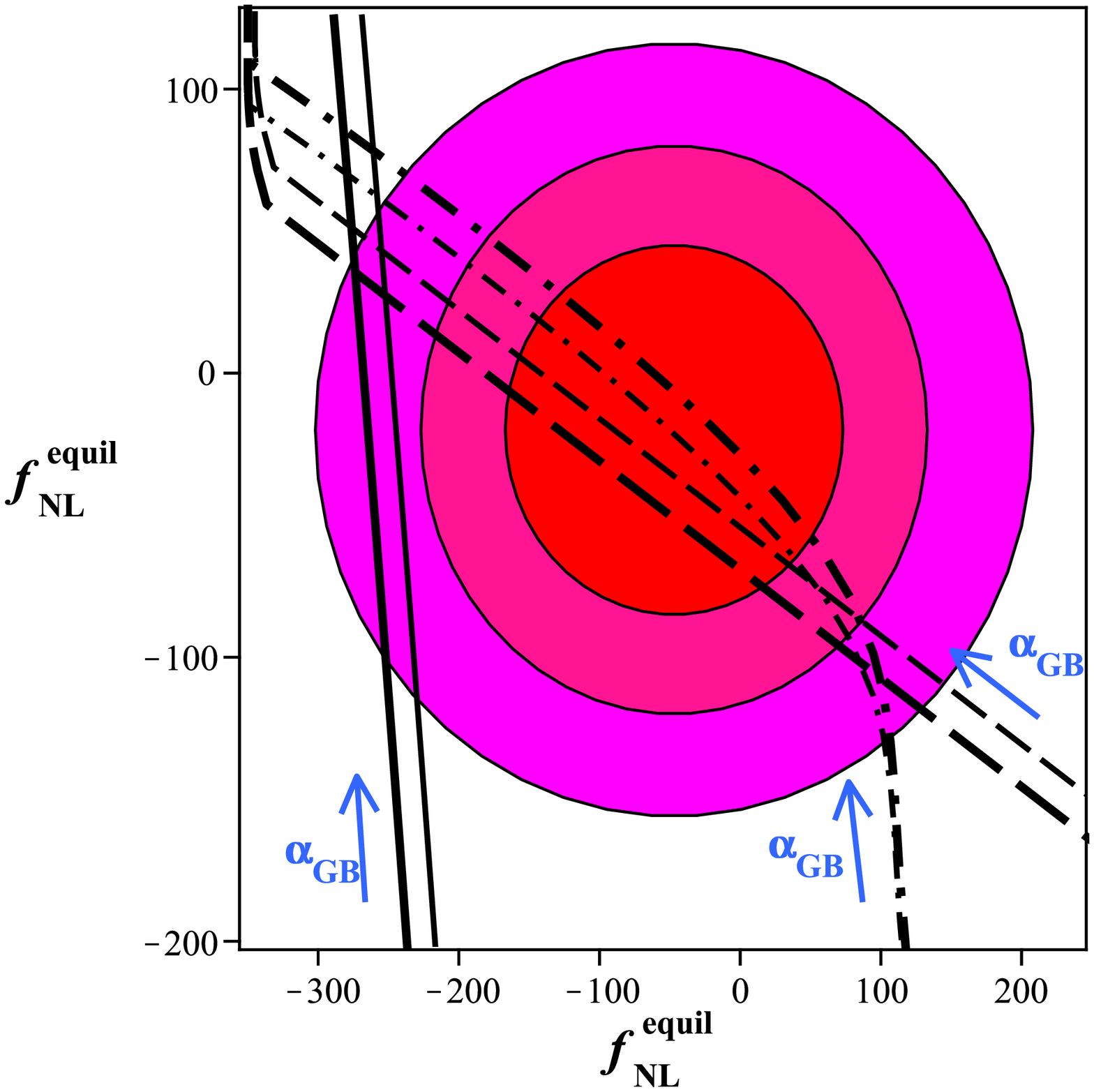} \vspace{6.5cm}
\end{center}
 \caption{\small {Evolution of the tensor to scalar ratio versus the
spectral index (left panel) and the amplitude of the non-Gaussianity
in the orthogonal configuration versus the equilateral
configuration(right panel), for the case with $f(\phi)=\beta
e^{\kappa\phi}$ and with an exponential potential, in the background
of the Planck+WMAP9+BAO data.}}
\end{figure}

If we take ${\cal{V}}(\phi)\sim \phi^{4}$, we find the following
expression for the number of e-fold parameter
\begin{equation}
N=\phi-\frac{16}{3}{\frac{\sigma\alpha_{_{GB}} \left(
6+6\kappa\phi+3\kappa^{2}\phi^{2}+\kappa^{3}\phi^{3}\right) }{{ e^{
\kappa\phi}}}}+{\frac{2\phi}{\sigma\beta}}-{\frac { 4\alpha_{_{GB}}
R_{_{GB}} \left(
\kappa^{3}\phi^{3}-3\kappa^{2}\phi^{2}+6\kappa\phi-6 \right) }{{
e^{-\kappa\phi}}\kappa^{4}\sigma}}\,\Bigg|_{hc}^{f}\,,\label{101}
\end{equation}
With this type of ${\cal{V}}$, we plot $r$ versus $n_{s}$ (the
dot-dashed lines in the left panel of figure 6). With
${\cal{V}}(\phi)\sim \phi^{4}$ and for $N=50$, the model lies inside
the joint $95\%$ CL of the Planck+WMAP9+BAO data if $8.06\times
10^{-5}<\alpha_{_{GB}}<7.4\times 10^{-3}$. Also, for $N=60$ the
model is compatible with observation if $8\times
10^{-5}<\alpha_{_{GB}}<7.33\times 10^{-3}$. We also plot the
evolution of $f^{ortho}$ versus $f^{equil}$ for ${\cal{V}}(\phi)\sim
\phi^{4}$ (the dot-dashed lines in the right panel of figure 6). We
find that this model lies inside the $95\%$ CL of the
Planck+WMAP9+BAO data if $8.11\times
10^{-5}<\alpha_{_{GB}}<7.69\times 10^{-3}$ for $N=50$, and
$8.06\times 10^{-5}<\alpha_{_{GB}}<7.8\times 10^{-3}$ for $N=60$.
Also, it is compatible with the $99\%$ CL of the Planck+WMAP9+BAO
data if $8.03\times 10^{-5}<\alpha_{_{GB}}<7.76\times 10^{-3}$ for
$N=50$, and $8\times 10^{-5}<\alpha_{_{GB}}<7.87\times 10^{-3}$ for
$N=60$.

Finally we solve the integral of equation (\ref{21}) with
${\cal{V}}(\phi)\sim e^{-\kappa\phi}$ and find
\begin{equation}
N=-\frac{1}{2}{\frac {\sigma\beta\ln  \left( -3\left( { e^{
\kappa\phi}} \right) ^{2}\beta+4\kappa^{4}\alpha_{_{GB}}
\sigma\beta+8\kappa^{4}\alpha_{_{GB}}+4\kappa^{4} \alpha_{_{GB}}^{2}
R_{_{GB}}\beta\right) }{\sigma\beta+2+ \alpha_{_{GB}}
R_{_{GB}}\beta}}\Bigg|_{hc}^{f}\,,\label{102}
\end{equation}
The solid lines in the left panel of figure 6 show the tensor to
scalar ratio versus the spectral index for ${\cal{V}}(\phi)\sim
e^{-\kappa\phi}$. In this case the model lies inside the joint
$95\%$ CL of the Planck+WMAP9+BAO data for $N=50$ and $N=60$, if
$3.73\times 10^{-4}<\alpha_{_{GB}}<1.24\times 10^{-2}$ and
$3.69\times 10^{-4}<\alpha_{_{GB}}<1.21\times 10^{-2}$ respectively.
Also, the solid lines in the left panel of figure 6 show the
evolution of $f^{ortho}$ versus $f^{equil}$ with
${\cal{V}}(\phi)\sim e^{-\kappa\phi}$. This model with an
exponential potential and with ${\cal{V}}(\phi)\sim e^{-\kappa\phi}$
is outside the joint $95\%$ CL of the Planck+WMAP9+BAO data, both
for $N=50$ and $N=60$. But, a comparison with the joint $99\%$ CL of
the Planck+WMAP9+BAO data shows that this model is compatible with
observation if $4.3\times 10^{-4}<\alpha_{_{GB}}<1.16\times 10^{-2}$
for $N=50$ and $4.24\times 10^{-4}<\alpha_{_{GB}}<1.3\times 10^{-2}$
for $N=60$. The results of these arguments are summarized in tables
3 and 4.

\begin{table*}
\begin{tiny}
\caption{\label{tab:3} The ranges of $\alpha_{_{GB}}$ for which the
values of the inflationary parameters $r$ and $n_{s}$ are compatible
with the joint $95\%$ CL of the Planck+WMAP9+BAO data with $f=\beta
e^{\kappa\phi}$.}
\begin{tabular}{ccccc}
\\ \hline \hline$V$& $N$ &${\cal{V}}(\phi)\sim \phi^{2}$&
${\cal{V}}(\phi)\sim \phi^{4}$&${\cal{V}}(\phi)\sim e^{-\kappa\phi}$\\ \hline\\
$\frac{\sigma}{2}\phi^{2}$& $N=50$ & $4.63\times
10^{-4}<\alpha_{_{GB}}<6.54\times 10^{-3}$ & $8.11\times
10^{-5}<\alpha_{_{GB}}< 3.42\times 10^{-2}$ & $1.3\times
10^{-4}<\alpha_{_{GB}}<9.03\times 10^{-3}$ \\\\
$\frac{\sigma}{2}\phi^{2}$& $N=60$ & $4.47\times
10^{-4}<\alpha_{_{GB}}<6.4\times 10^{-3}$ & $8\times
10^{-5}<\alpha_{_{GB}}<3.38\times 10^{-2}$ &
$1.41\times 10^{-4}<\alpha_{_{GB}}<9.1\times 10^{-3}$ \\\\
$\frac{\sigma}{4}\phi^{4}$& $N=50$ & $3\times
10^{-3}<\alpha_{_{GB}}<4.5\times 10^{-2}$  & $0.9\times
10^{-4}<\alpha_{_{GB}}<8.5\times 10^{-3}$  & $6.86\times
10^{-4}<\alpha_{_{GB}}<6.44\times 10^{-3}$\\\\
$\frac{\sigma}{4}\phi^{4}$& $N=60$ &  $3.08\times
10^{-3}<\alpha_{_{GB}}<4.44\times 10^{-2}$ & $0.83\times
10^{-4}<\alpha_{_{GB}}<8.38\times
10^{-3}$ & $6.67\times 10^{-4}<\alpha_{_{GB}}<6.31\times 10^{-3}$\\\\
$\sigma e^{-\kappa\phi}$& $N=50$ & $6.1\times
10^{-4}<\alpha_{_{GB}}<8.9\times 10^{-3}$ & $8.06\times
10^{-5}<\alpha_{_{GB}}<7.4\times 10^{-3}$ & $3.73\times
10^{-4}<\alpha_{_{GB}}<1.24\times
10^{-2}$ \\\\
$\sigma e^{-\kappa\phi}$& $N=60$ & $5.97\times
10^{-4}<\alpha_{_{GB}}<8.81\times 10^{-3}$ & $8\times
10^{-5}<\alpha_{_{GB}}<7.33\times 10^{-3}$ & $3.69\times 10^{-4}<\alpha_{_{GB}}<1.21\times 10^{-2}$ \\ \hline\\\\
\end{tabular}
\end{tiny}
\end{table*}

\begin{table*}
\begin{tiny}
\caption{\label{tab:4} The ranges of $\alpha_{_{GB}}$ for which the
values of the inflationary parameters $f^{ortho}$ and $f^{equil}$
are compatible with the joint $99\%$ CL of the Planck+WMAP9+BAO data
with $f=\beta e^{\kappa\phi}$.}
\begin{tabular}{ccccc}
\\ \hline \hline$V$& $N$ &${\cal{V}}(\phi)\sim \phi^{2}$&
${\cal{V}}(\phi)\sim \phi^{4}$&${\cal{V}}(\phi)\sim e^{-\kappa\phi}$\\ \hline\\
$\frac{\sigma}{2}\phi^{2}$& $N=50$ & $4.66\times
10^{-4}<\alpha_{_{GB}}\leq6.774\times 10^{-3}$ & $3.1\times
10^{-4}<\alpha_{_{GB}}<
4.02\times 10^{-3}$ & $1.56\times 10^{-4}<\alpha_{_{GB}}<8.9\times 10^{-3}$ \\\\
$\frac{\sigma}{2}\phi^{2}$& $N=60$ & $4.41\times
10^{-4}<\alpha_{_{GB}}<6.73\times 10^{-3}$ & $3.18\times
10^{-4}<\alpha_{_{GB}}< 4.15\times 10^{-3}$ & $1.62\times 10^{-4}<\alpha_{_{GB}}<9\times 10^{-3}$ \\\\
$\frac{\sigma}{4}\phi^{4}$& $N=50$ & $1.3\times
10^{-3}<\alpha_{_{GB}}<4.26\times 10^{-2}$  & $1.43\times
10^{-4}<\alpha_{_{GB}}<8.02\times 10^{-3}$ & $6.62\times
10^{-4}<\alpha_{_{GB}}<8.3\times 10^{-3}$ \\\\
$\frac{\sigma}{4}\phi^{4}$& $N=60$ &  $1\times
10^{-3}<\alpha_{_{GB}}<4.38\times 10^{-2}$ & $1.15\times
10^{-4}<\alpha_{_{GB}}<8.26 \times 10^{-3}$ & $6.6\times
10^{-4}<\alpha_{_{GB}}\leq6.265\times 10^{-3}$\\\\
$\sigma e^{-\kappa\phi}$& $N=50$ & $6.23\times
10^{-4}<\alpha_{_{GB}}<9.36\times 10^{-3}$ & $8.03\times
10^{-5}<\alpha_{_{GB}}<7.76\times 10^{-3}$ & $4.3\times
10^{-4}<\alpha_{_{GB}}<1.16\times 10^{-2}$ \\\\
$\sigma e^{-\kappa\phi}$& $N=60$ & $6.125\times
10^{-4}\leq\alpha_{_{GB}}\leq9.141\times 10^{-3}$ & $8\times
10^{-5}<\alpha_{_{GB}}<7.87\times 10^{-3}$ &
$4.24\times 10^{-4}<\alpha_{_{GB}}<1.3\times 10^{-2}$ \\ \hline\\\\
\end{tabular}
\end{tiny}
\end{table*}

\section{Conclusion}
In this paper, we have considered a DBI model which is non-minimally
coupled to a Gauss-Bonnet term. We have studied the cosmological
dynamics of this model in the early times of the Universe history.
We have calculated the inflationary parameters and the primordial
density perturbations with details. If the DBI field is the only
field in the inflation period as responsible for the inflation, the
perturbations are adiabatic. But, in this paper since the DBI field
interacts with the Gauss-Bonnet term, the isocurvature perturbations
can be generated. Also, because of the non-minimal coupling between
the DBI field and the Gauss-Bonnet term, the two metric
perturbations are different. The non-Gaussianity of the primordial
density perturbations in this model has been explored by studying
the three point correlators. To this end, we have expanded the
action up to the cubic order in the small fluctuations around the
homogeneous background solution. Then, by using the interaction
picture we have calculated the three point correlation functions. In
the three point correlators, there are functions which depend on the
momenta and are dubbed shape of non-Gaussianity. The momenta form a
triangle and every shape has a pick in a configuration of triangle.
In this work we have focused on the equilateral and orthogonal
shapes which have a pick in $k_{1}=k_{2}=k_{3}$. We have expressed
the leading-order bispectrum in terms of the equilateral basis
$S_{*}^{equil}$ and the orthogonal basis $S_{*}^{ortho}$ and have
found the amplitudes of the non-Gaussianity in the equilateral and
orthogonal configurations. After obtaining the main equations, we
have compared our setup with the recent observational data. We have
taken the Gauss-Bonnet coupling as
$\alpha(\phi)=\alpha_{_{GB}}{\cal{V}}$ and have focused on the
values of $\alpha_{_{GB}}$. We have considered two functions for
$f(\phi)$ as $f=\beta\phi^{-4}$ and $f=\beta e^{\kappa\phi}$. For
every $f$, we have adopted three potentials as
$V(\phi)=\frac{\sigma}{2}\phi^{2}$, $\frac{\sigma}{4}\phi^{4}$ and
$\sigma e^{-\kappa\phi}$ and for every potential we have considered
three functions for ${\cal{V}}$ as ${\cal{V}}\sim \phi^{2}$,
$\phi^{4}$ and $e^{\kappa\phi}$. By choosing these functions we have
studied the evolution of the tensor to scalar ratio versus the
scalar spectral index and the amplitude of the non-Gaussianity in
the orthogonal configuration versus the equilateral configuration in
the background of the joint Planck+WMAP9+BAO data. Also, we have
obtained some constraints on the Gauss-Bonnet coupling term,
$\alpha_{_{GB}}$ that are summarized in tables. Our study shows that
for $f=\beta\phi^{-4}$, the amplitude of the non-Gaussianity of the
primordial perturbation in the model with a quadratic or exponential
potentials and ${\cal{V}}\sim \phi^{4}$ lies outside the joint
$95\%$ CL of the Planck+WMAP9+BAO data. But, in some ranges of the
Gauss-Bonnet coupling term, it is inside the joint $99\%$ CL of the
Planck+WMAP9+BAO data. Also, for $f=\beta e^{\kappa\phi}$ the
amplitude of the non-Gaussianity of the model with a quadratic
potential and ${\cal{V}}\sim \phi^{4}$ and with an exponential
potential and ${\cal{V}}\sim e^{-\kappa\phi}$ is outside the joint
$95\%$ CL of the Planck+WMAP9+BAO data but inside the joint $99\%$
CL of the Planck+WMAP9+BAO data in some ranges of $\alpha_{_{GB}}$.
In other cases we have found the range of $\alpha_{_{GB}}$ in which
this setup is compatible with the observational data. Although our
tables contain a variety of domains for $\alpha_{_{GB}}$, but
inspection of these ranges show that the constraint $1.14\times
10^{-5}\leq\alpha_{_{GB}}<4.5\times 10^{-2}$ contains the most
general domain for $\alpha_{_{GB}}$. In summary, in confrontation
with recent data, the Gauss-Bonnet coupling $\alpha_{_{GB}}$ is
restricted to this domain.\\\\

\textbf{Appendix A}
\begin{eqnarray}
S_{3}=\int dt\, d^{3}x\, a^{3}\,
\Bigg\{\bigg[\frac{3H}{\kappa^{2}}+\frac{\dot{\phi}^{2}}{2\sqrt{1-f\dot{\phi}^{2}}}-\frac{f\,
\dot{\phi}^{4}}{\big(1-f\dot{\phi}^{2}\big)^{\frac{3}{2}}}+\frac{f^{2}\dot{\phi}^{6}}
{2\big(1-f\dot{\phi}^{2}\big)^{\frac{5}{2}}}-80H^{3}\dot{\alpha}\bigg]\Phi^{3}\nonumber
\end{eqnarray}
\begin{eqnarray}
+\bigg[\Big(144H^{3}\dot{\alpha}
-\frac{9H^{2}}{\kappa^{2}}+\frac{3\dot{\phi}^{2}}{2\sqrt{1-f\dot{\phi}^{2}}}+
\frac{3f\,\dot{\phi}^{4}}{2\big(1-f\dot{\phi}^{2}\big)^{\frac{3}{2}}}\Big)\Psi-\Big(\frac{6H}{\kappa^{2}}
-48H^{2}\dot{\alpha}\Big)\dot{\Psi}-\frac{16H\dot{\alpha}}{a^{2}}\partial^{2}\Psi\nonumber
\end{eqnarray}
\begin{eqnarray}
+\frac{2\kappa^{-2}H+16H^{2}\dot{\alpha}}{a^{2}}\partial^{2}B\bigg]\Phi^{2}
+\bigg[\Big(\frac{18H}{\kappa^{2}}+216H^{2}\dot{\alpha}\Big)\dot{\Psi}\Psi
+\frac{16\dot{\alpha}}{a^{2}}\dot{\Psi}\partial^{2}\Psi+\frac{\Big(16H\dot{\alpha}-\frac{2}{\kappa^{2}}\Big)}
{a^{2}}\Psi\partial^{2}\Psi\nonumber
\end{eqnarray}
\begin{eqnarray}
-\frac{2\kappa^{-2}H+16H^{2}\dot{\alpha}}{a^{2}}\partial_{i}\Psi\partial_{i}B
+\frac{\Big(8H\dot{\alpha}-\kappa^{-1}\Big)}{a^{2}}\Big(\partial\Psi\Big)^{2}+\frac{12H\dot{\alpha}-\frac{1}{2\kappa^{2}}}{a^{4}}
\big(\partial_{i}\partial_{j} B\,\partial_{i}\partial_{j}
B-\partial^{2}B\partial^{2}B\big)\nonumber
\end{eqnarray}
\begin{eqnarray}
+\frac{8\dot{\alpha}}{a^{4}} \big(\partial_{i}\partial_{j}
B\,\partial_{i}\partial_{j}
\Psi-\partial^{2}B\partial^{2}\Psi\big)-\frac{2\kappa^{-2}H+24H^{2}\dot{\alpha}}{a^{4}}\Psi\partial^{2}
B+\frac{48H\dot{\alpha}-2\kappa^{-2}}{a^{2}}\dot{\Psi}\partial^{2}B\nonumber
\end{eqnarray}
\begin{eqnarray}
+\Big(3\kappa^{-2}-72H\dot{\alpha}\Big)\dot{\Psi}^{2}\bigg]\Phi+8\dot{\alpha}\dot{\Psi}^{3}+\frac{\kappa^{-2}-8\ddot{\alpha}}
{a^{2}}\Psi\Big(\partial\Psi\Big)^{2}+\Big(72H\dot{\alpha}-9\kappa^{-2}\Big)\dot{\Psi}^{2}\,\Psi\nonumber
\end{eqnarray}
\begin{eqnarray}
+\frac{2\kappa^{-2}-16H\dot{\alpha}}{a^{2}}\dot{\Psi}\partial_{i}\Psi\partial_{i}B
-\frac{8\dot{\alpha}}{a^{2}}\dot{\Psi}^{2}\partial^{2}B
+\frac{2\kappa^{-2}-16H\dot{\alpha}}{a^{2}}\dot{\Psi}\Psi
\partial^{2}B -\frac{2\kappa^{-2}-16H\dot{\alpha}}{a^{4}} \partial_{i}\Psi \partial_{i} B \partial^{2} B \nonumber
\end{eqnarray}
\begin{eqnarray}
+\frac{(\frac{3}{2\kappa^{2}}-12H\dot{\alpha})\Psi
-4\dot{\alpha}\dot{\Psi}}{a^{4}}\big(\partial_{i}\partial{j}B\partial_{i}\partial_{j}B
-\partial^{2}B\partial^{2}B\big)\Bigg\}\,.
\end{eqnarray}


\begin{thebibliography}{100}
\bibitem{Gut81} A. Guth, Phys. Rev. D, \textbf{23}, 347 (1981).

\bibitem{Lin82} A. D. Linde, Phys. Lett. , \textbf{108 B}, 389
(1982)

\bibitem{Alb82} A. Albrecht and P. Steinhard, Phys. Rev. D, \textbf{48}, 1220
(1982).

\bibitem{Lin90} A. D. Linde, \emph{Particle Physics and Inflationary Cosmology}
(Harwood Academic Publishers, Chur, Switzerland, 1990).
[arXiv:hep-th/0503203].

\bibitem{Lid00a} A. Liddle and D. Lyth, \emph{Cosmological Inflation and Large-Scale
Structure}, (Cambridge University Press, 2000).

\bibitem{Lid97} J. E. Lidsey et al, Abney, Rev. Mod. Phys.,
\textbf{69}, 373, (1997).

\bibitem{Rio02} A. Riotto, [arXiv:hep-ph/0210162].

\bibitem{Lyt09} D. H. Lyth and A. R. Liddle, \emph{The Primordial Density
Perturbation} (Cambridge University Press, 2009).

\bibitem{Mal03} J. M. Maldacena, JHEP, \textbf{0305},
013, (2003).

\bibitem{Bar04} N. Bartolo, E. Komatsu, S. Matarrese and A. Riotto, Phys.Rept.,
 \textbf{402}, 103, (2004).

\bibitem{Che10} X. Chen, Adv. Astron. \textbf{2010}, 638979, (2010).

\bibitem{Fel11a} A. De Felice and S. Tsujikawa, Phys. Rev. D, \textbf{84},
083504, (2011).

\bibitem{Fel11b} A. De Felice and S. Tsujikawa, JCAP, \textbf{1104},
029, (2011).

\bibitem{Bab04a} D. Babich, P. Creminelli and M. Zaldarriaga, JCAP, \textbf{0408},
009, (2004).

\bibitem{Che08} C. Cheung, P. Creminelli, A. L. Fitzpatrick, J. Kaplan and L.
Senatore, JHEP, \textbf{0803}, 014, (2008).

\bibitem{Wan03} Y. Wang, [arXiv:1303.1523 [hep-th]].

\bibitem{Lan11} D. Langlois, [arXiv:1102.5052 [astro-ph.CO]].

\bibitem{Kom05} E. Komatsu, D. N. Spergel and B. D. Wandelt, Astrophys. J. \textbf{634},
14, (2005).

\bibitem{Cre06} P. Creminelli, A. Nicolis, L. Senatore, M. Tegmark and M.
Zaldarriaga, JCAP, \textbf{0605}, 004, (2006).

\bibitem{Lig06} M. Liguori, F. K. Hansen, E. Komatsu, S. Matarrese and A.
Riotto, Phys. Rev. D, \textbf{73}, 043505, (2006).

\bibitem{Yad07} A. P. S. Yadav, E. Komatsu and B. D. Wandelt, Astrophys. J., \textbf{664},
680,(2007).

\bibitem{Zwi85} B. Zwiebach, Phys. Lett. B, \textbf{156}, 315, (1985).

\bibitem{Bou85} D. G. Boulware and S. Deser, Phys. Rev. Lett., \textbf{55}, 2656, (1985).

\bibitem{Noj05} S. Nojiri, S. D. Odintsov and M. Sasaki, Phys. Rev. D, \textbf{71},
123509, (2005).

\bibitem{Noj07} S. Nojiri, S. D. Odintsov and P. V. Tretyakov, Phys. Lett. B, \textbf{651}, 224,
(2007).

\bibitem{Guo09} Z. K. Guo and D. J. Schwarz, Phys. Rev. D, \textbf{80}, 063523,
(2009).

\bibitem{Guo10} Z. K. Guo and D. J. Schwarz, Phys. Rev. D, \textbf{81}, 123520,
(2010).

\bibitem{Bro07} R. A. Brown, \emph{Brane world cosmology with Gauss-Bonnet and induced gravity terms},
(PhD Thesis, 2007), [arXiv:gr-qc/0701083].

\bibitem{Bam07} K. Bamba, Z. K. Guo and N. Ohta, Prog. Theor. Phys., \textbf{118},
879, (2007).

\bibitem{And07} K. Andrew, B. Bolen and C. A. Middleton, Gen. Rel. Grav., \textbf{39},
2061, (2007).

\bibitem{Noz08} K. Nozari and B. Fazlpour, JCAP, \textbf{0806}, 032,
(2008).

\bibitem {Noz09a} K. Nozari, and N. Rashidi, Int. J. Thoer. Phys., \textbf{48},
2800, (2009).

\bibitem {Noz09b} K. Nozari, and N. Rashidi, JCAP, \textbf{0909},
014, (2009).

\bibitem {Noz09c} K. Nozari, and N. Rashidi, Int. J. Mod. Phys. D, \textbf{19},
219, (2009).

\bibitem{Sil04} E. Silverstein and D. Tong, Phys. Rev. D, \textbf{70},
103505, (2004).

\bibitem{Ali04} M. Alishahiha, E. Silverstein, and D. Tong, Phys.
Rev. D, \textbf{70}, 123505, (2004).

\bibitem{Hua06} M. x. Huang and G. Shiu, Phys. Rev. D, \textbf{74}, 121301, (2006).

\bibitem{Che07} X. Chen, M. x. Huang, S. Kachru and G. Shiu, JCAP, \textbf{0701},
002, (2007).

\bibitem{Ade13a} P. A. R. Ade et al., [arXiv:1303.5082].

\bibitem{Ade13b} P. A. R. Ade et al., [arXiv:1303.5084].

\bibitem{Bar80} J. Bardeen, PRD, \textbf{22}, 1882, (1980).

\bibitem{Muk92} V. F. Mukhanov, H. A. Feldman and R. H. Brandenberger, Phys. Rept.,
\textbf{215}, 203, (1992).

\bibitem{Ber95} E. Bertschinger, [arXiv:astro-ph/9503125], (1995).

\bibitem{Bas99} B. A. Bassett, F. Tamburini, D. I. Kaiser and R. Maartens, Nucl.
Phys. \textbf{B}, \textbf{561}, 188, (1999).

\bibitem{Gor01} C. Gordon, D. Wands, B. A. Basset and R. Maartens, PRD, \textbf{63},
023506, (2001).

\bibitem {Lop04} M. Bouhamdi-Lopez, R. Maartens and D. Wands, PRD, \textbf{70},
123519, (2004).

\bibitem{Kal05} N. Kaloper, PRD, \textbf{71}, 086003, (2005).

\bibitem{Maa00} R. Maartens, D. Wands, B. A. Bassett and I. Heard, PRD,
\textbf{62}, 041301, (2000).

\bibitem{Lan00} D. Langlois, R. Maartens and D. Wands, Phys. Lett. \textbf{B},
\textbf{489}, 259, (2000).

\bibitem{Lan07} D. Langlois and F. Vernizzi, JCAP, \textbf{02}, 017, (2007).

\bibitem{Bar83} J. M. Bardeen, P. J. Steinhardt and M. S. Turner, PRD, \textbf{28},
679, (1983).

\bibitem{Wan00} D. Wands, K. A. Malik, D. H. Lyth and A. R. Liddle, PRD,
\textbf{62}, 043527, (2000).

\bibitem{Amn06} L. Amendola, C. Charmousis and S. C. Davis, JCAP \textbf{0612}, 020,
(2006).

\bibitem{Amn07} L. Amendola, C. Charmousis and S. C. Davis, JCAP \textbf{0710}, 004,
(2007).

\bibitem{Noz12} K. Nozari and N. Rashidi, Phys. Rev. D, \textbf{86}, 043505
(2012).

\bibitem{Noz13} K. Nozari and N. Rashidi, Phys. Rev. D, \textbf{88}, 023519
(2013).

\bibitem{Arn60} R. L. Arnowitt, S. Deser and C. W. Misner, Phys. Rev., \textbf{117},
1595, (1960).

\bibitem{Koy10} K. Koyama, Class. Quant. Grav., \textbf{27}, 124001 (2010).

\bibitem{Lyt05} D. Lyth and Y. Rodriguez, Phys.Rev. D, \textbf{71}, 123508, (2005).

\bibitem{See05} D. Seery and J. E. Lidsey, JCAP, \textbf{0506}, 003,
(2005).

\bibitem{Che02} X. Chen, M. x. Huang, S. Kachru and G. Shiu, JCAP, \textbf{0701},
002, (2007).

\bibitem{Gan94} A. Gangui, F. Lucchin, S. Matarrese and S. Mollerach, ApJ, \textbf{430},
447,(1994).

\bibitem{Ver00} L. Verde, L. Wang, A. F. Heavens and M. Kamionkowski, MNRAS,
\textbf{313}, 141, (2000).

\bibitem{Wan00b} L. Wang and M. Kamionkowski, Phys. Rev. D, \textbf{61},
063504, (2000).

\bibitem{Kom01} E. Komatsu and D. N. Spergel, Phys. Rev. D, \textbf{63},
063002, (2001).

\bibitem{Bab04b} D. Babich, P. Creminelli and M. Zaldarriaga, J. Cosmology
Astropart. Phys., \textbf{8}, 9, (2004).

\bibitem{Sen10} L. Senatore, K. M. Smith and M. Zaldarriaga, J. Cosmology
Astropart. Phys., \textbf{1}, 28,  (2010).

\bibitem{Fel13} A. De Felice, S. Tsujikawa, JCAP, \textbf{03}, 030,
(2013).

\bibitem{Tsu13} S. Tsujikawa, J. Ohashi, S. Kuroyanagi and A. De
Felice, Phys.Rev.D, \textbf{88}, 023529, (2013).





\end{thebibliography}
\end{document}